%% file: paper_revmap.tex
\documentclass{emulateapj}

\usepackage{amsmath,amsfonts, amssymb}

\usepackage{mathrsfs}

\usepackage{dsfont}

\usepackage{subfigure}
\usepackage{url}
\usepackage{paralist}
\newcommand\ionscale[2]{#1$\;${\scshape{#2}}} 

\shorttitle{Estimating Black Hole Masses in Hundreds of Quasars}
\shortauthors{Hernitschek et al.}

\begin{document}


\title{Estimating Black Hole Masses in Hundreds of Quasars}


\author{Nina Hernitschek and Hans-Walter Rix}
\affil{Max-Planck-Institut f{\"u}r Astronomie, K{\"o}nigstuhl 17, D-69117 Heidelberg}
\email{hernitschek@mpia-hd.mpg.de}

\author{Jo Bovy\altaffilmark{1}}      
\affil{Institute for Advanced Study, School of Natural Sciences, Einstein Drive, Princeton, NJ 08540}

\and

\author{Eric Morganson}
\affil{Harvard-Smithsonian Center for Astrophysics, 60 Garden St, Cambridge, MA 02138 }

\altaffiltext{1}{Hubble Fellow}



\begin{abstract}
We explore the practical feasibility of AGN broad-band reverberation mapping and present first results. We lay out and apply a rigorous approach for stochastic reverberation mapping of unevenly sampled multi-broad-band flux measurements, assuming that the broad-line region (BLR) line flux is contributing up to 15 \% in some bands, and is directly constrained by one spectroscopical epoch. 

The approach describes variations of the observed flux as the continuum, modeled as a stochastic Gaussian process, and emission line contribution, modeled as a scaled, smoothed and delayed version of the continuum. This approach is capable not only to interpolate in time between measurements, but also to determine confidence limits on continuum -- line emission delays. This approach is applied to SDSS observations in 'Stripe 82' (S82) providing flux measurements precise to 2 \% at $\sim$ 60 epochs over $\sim$ 10 years. The strong annual variations in the epoch sampling prove a serious limitation in practice. Also, suitable redshift ranges must be identified, where strong broad emission line contribute to one filter, but not to another. 

Through generating and evaluating problem-specific mock data, we verify that S82-like data can constrain $\tau_{\mathrm{delay}}$ for a simple transfer function model. In application to real data, we estimate $\tau_{\mathrm{delay}}$ for 323 AGN with $0.225 < z < 0.846$, combining information for different objects through the ensemble-scaling relationships for BLR size and BH mass. Our analysis tentatively indicates a 1.7 times larger BLR size of H$\alpha$ and \ionscale{Mg}{ii} compared to \cite{Kaspi2000} and \cite{Vestergaard2002}, but the seasonal data sampling casts doubt on the robustness of the inference.
\end{abstract}




\input{Introduction}

\input{section2} 
\input{section3}

\input{section4}

\input{section5}

\input{section6} 
\input{section7}

\section{Acknowledgments}

The research leading to these results has received
funding from the European Research Council
under the European Union's Seventh Framework
Programme (FP 7) ERC Grant Agreement n.
[321035].

Jo Bovy was supported by
NASA through Hubble Fellowship grant HST-HF-51285.01 from the Space
Telescope Science Institute, which is operated by the Association of
Universities for Research in Astronomy, Incorporated, under NASA contract
NAS5-26555.

\clearpage
\appendix
    \input{Appendix}

\clearpage
   
%



 \clearpage
 
  \input{Tables}

\end{document}

%% file: Introduction.tex
\section{INTRODUCTION}
\label{sec:Introduction}

Quasars have long been known to exhibit quite rapid optical variability that can be attributed to variations in the luminosity of the accretion disk surrounding a black hole of typically 10$^8$ M$_{\sun}$ \citep{Smith1963, Greenstein1964}.

Reverberation mapping \citep{Blandford1982,Kaspi2000} is an established way for estimating the size of the AGN's broad line region (BLR). The continuum radiation from the accretion disk photo-ionizes and excites gas clouds close to the black hole to produce broad (about 1,000 to 10,000 km/s) emission lines. In reverberation mapping, the time delay $\tau_{\mathrm{delay}}$ between observed variations in the accretion disk continuum and in the broad emission lines is a proxy for $R_{\mathrm{BLR}}$ as light-travel time arguments lead to $R_{\mathrm{BLR}} \propto c \, \tau_{\mathrm{delay}}$. For Keplerian motions of the BLR clouds, this implies for the mass of the central black hole, $M_{\mathrm{BH}}$
\begin{equation}
M_{\mathrm{BH}} = f\, \frac{\Delta V^2 c  \, \tau_{\mathrm{delay}}}{G}
\label{eqn:massrev}
\end{equation}
where $G$ is the gravitational constant and $f$ is a proportionality factor of order unity that depends on the geometry and kinematics of the BLR (e.g., \citealt{Peterson1999}).
Detailed descriptions of this method as well as applications can be found e.g. in \citet{PetersonBook}, \citet{Peterson1999}, \citet{Peterson2013}, \citet{Kaspi2000}.

With the benefits of many spectral observational epochs in reverberation mapping campaigns, giving reliable average emission line widths $\Delta v$, reverberation mapping provides reliable direct measurements of the size of the BLR and the black hole mass (e.g. \citealt{Kaspi2007}, \citealt{Peterson2004}). \citet{Kaspi2000} presents a $M_{\mathrm{BH}}$ - $L$ relation based on spectrophotometrical reverberation measurements for a sample of 17 Palomar-Green quasars, and a total of 34 sources, including low-$L$ AGN.

They obtained the size of their BLRs and determined relationships between line luminosities, BLR sizes and central black hole masses to find that the BLR size scales with the rest-frame 5100 $\mathrm{\mathring{A}}$ luminosity as
\begin{equation}
R_{\mathrm{BLR}}= \left(  32.0  ^{+2.0}_{-1.9} \right) \left( \frac{\lambda L_{\lambda}(5100 \; \mathrm{\mathring{A}})}{10^{44} \mathrm{ \;erg \;s^ {-1}}} \right) ^{0.700 \pm 0.033} \textrm{light days}.
\end{equation}
Comparable studies have been done by Vestergaard et al. \citep{Vestergaard2002, Vestergaard2006} and \citet{Bentz2009}. They found four empirical mass scaling relationships between line widths and luminosity for estimating $M_{\mathrm{BH}}$ in nearby AGNs and distant luminous quasars. Those mass estimates are quite rough. For example, \citet{Vestergaard2002} indicate the absolute uncertainties in masses estimated from the relationships of a factor of $\sim$4.

The purpose of the present study is to estimate BLR sizes from existing multi-epoch broad band flux measurements. We build on the formalism from \cite{Zu2011} and \cite{Chelouche2012} and extend it to handle sparsely sampled broad-band photometric data. 

The basic idea is that some of the photometric passbands contain only continuum emission, while in other passbands, there are significant contributions from
continuum and (temporally delayed) line flux.
Recent results of \citet{Haas2011}, who tested narrow-band photometric reverberation
mapping, and \cite{Chelouche2012} and \cite{CheloucheZucker2013}, who suggested broad-band photometric
reverberation mapping, illustrate the potential of using photometric data. \citet{Zu2013a}
have compared results of spectroscopic and photometric reverberation mapping applied to
the Palomar-Green quasars and OGLE-III and IV. They have found that the photometric
approach is capable of compete with spectroscopic reverberation mapping if very small
photometric uncertainties are available and strong lines (H$\alpha$, H$\beta$) are used. Upcoming photometric
surveys such as the LSST are planned to continuously monitor at least 10$^7$ quasars
(0 $< z <$ 6) over the next decade \citep{MacLeod2012}. Broad band photometric reverberation
mapping can utilize such data to make the mass estimate of large samples of objects feasible
to increase the number of reverberation mapped objects by several orders of magnitude, e.g. \cite{Chelouche2014}.

This formalism is first tested on mock light curves generated by a Gaussian stochastic process. Then it is applied to a suitable set of multi-band quasar light curves, drawn from the nearly 10,000 spectroscopically confirmed quasars in SDSS Stripe 82 \citep{Schmidt2010, Schneider2007}, which are complemented by a spectroscopic measurement of the emission line widths for each quasar at one epoch. 
Due to small expected signal and the S82 time sampling, we found
it useful to not focus on the $\tau_{\mathrm{delay}}$ estimates of individual objects, but to presume that there is a $R_{\mathrm{BLR}}(L)$ relation, and determine its scaling normalization in different redshift and luminosity regimes, by jointly modelling several light curves. 
The reverberation mapping results are compared to estimates from $M_{\mathrm{BH}}$ - $L$ relationships in \cite{Kaspi2000}, \cite{Vestergaard2002} and \cite{Bentz2013}.

This paper is organized as follows. In Section \ref{sec:ReverberationMapping} we give an overview of reverberation mapping, especially the theory behind this method, and preliminary reverberation mapping results. After introducing methods of describing quasar variability as a stochastic process in Section 3, we introduce the stochastic reverberation mapping approach in Section 4. This is outlined in more detail in the Appendix, where the mathematical framework of the stochastical process model for the light curve and the application of the method to data are described. After describing the application to SDSS S82 data in Section 5, results are shown in Section 6. In Section 7, we conclude with a discussion of results. In the Table Section, we provide a complete list of estimated masses for all those quasars from our samples where reverberation mapping was carried out successfully.

%% file: section2.tex
\section{REVERBERATION MAPPING}
\label{sec:ReverberationMapping}

Over the duration of a reverberation mapping program, the continuum behavior can be written as $f^c(t)= \langle f^c \rangle +\Delta f^c(t)$, where $\Delta f^c(t)$ is the continuum light curve relatively to its mean value $\langle f^c \rangle$ given in arbitrary flux units \citep{PetersonBook}. Integrated over the velocity-dependent line profile, the emission-line response can be written as a function of the line-of-sight velocity $v$ as $f^e(t)= \langle f^e \rangle + \Delta f^e(t)$. On reverberation timescales (weeks to years), both continuum and emission-line variations are usually rather small (typically $\sim$10--20 \%), so the response of the emission line flux to (e.g. increased) continuum flux can be modeled by a convolution integral \citep{PetersonBook}
\begin{equation}
\Delta f^e(t)= \int \Psi (\tau_{\mathrm{delay}})\Delta f^c(t-\tau_{\mathrm{delay}}) \,  \mathrm{d}\tau_{\mathrm{delay}},
\label{eqn:transferequation}
\end{equation}
which is usually known as the \textit{transfer equation}, where $\Psi (\tau_{\mathrm{delay}})$ is the \textit{transfer function}. In its mathematically simplest form this transfer function can be taken as a $\delta$ function that is offset in time by $\tau_{\mathrm{delay}}$, $\Psi (\tau_{\mathrm{delay}})=\delta(t-\tau_{\mathrm{delay}})$. The BLR geometry and detailed spectrosopic data for nearby objects indicate that such a transfer function is too simplistic; nonetheless, we will use this approach involving a $\delta$ function transfer function in this study, as broad-band reverberation mapping is unlikely to yield any velocity-dependent information. This leads to a scaling and a delay during the transfer function, whereas the usage of other transfer functions can also lead to a smoothing.

The goal of reverberation mapping is to use the observables, namely the continuum light curve $f^c(t)$ and the emission-line light curve $f^e(t)$, and invert the transfer equation \eqref{eqn:transferequation} in order to recover the velocity--delay map $\Psi(\tau_{\mathrm{delay}})$, or at least to make inferences about $\tau_{\mathrm{delay}}$ \citep{PetersonBook}.

When spectroscopic reverberation mapping data are available, a cross-correlation approach between the pure line and continuum light curves has often been employed \citep{PetersonBook}.
For the case of broad-band photometric light curve data, a simple model to illustrate the calculations based on photometric data is
\begin{align}\label{eqn:rgmodel}
\begin{split}
f_k(t) & =  f_k^c(t) \\
 f_l(t) &=   f_l^c(t) +  f_l^e(t) \\
  &= s \,f_k^c(t) + e \, f_k^c(t-\tau_{\mathrm{delay}}) 
\end{split}
\end{align}
where $f(t)$ is the flux on each time, index $k$ denotes a band with only continuum, $l$ a continuum and emission line contribution band, superscripts $^c$ and $^e$ denote continuum or emission line contributions, $\tau_{\mathrm{delay}}$ is the delayed response and $s$, $e$ are scaling factors. For estimating the delay $\tau_{\mathrm{delay}}$ between the continuum flux $f_k^c(t)$ and the emission line flux $f_l^e(t)$, one must compute the cross-correlation function (CCF) between these two components of the light curve \citep{Edri2012}:
\begin{eqnarray*}
\mathrm{CCF}(\Delta t)&=&f_l^e(t+\Delta t)*f_l^c(t)\\
&=&\left(f_l(t+\Delta t)-f_l^c(t+\Delta t)\right)*f_l^c(t)
\end{eqnarray*}
where $*$ denotes the integral over time (convolution between the two functions). 

The peak (maximum) of the $\mathrm{CCF}(\Delta t)$ gives the required time delay  $\tau_{d\mathrm{elay}}$. \ \\
We now assume the time variability of the continuum flux in the $l$ band is the same as in the $k$ band. This is a good approximation in the optical since the continuum is 75 \% to 95 \% of the total flux, where the remaining variable flux is mostly coming from the broad lines. With this approximation, $f_l^c(t) \approx f_k(t)$, and $e \ll 1$ (Equ. \eqref{eqn:rgmodel}), so that $\mathrm{CFF}(0) \sim 1$, this leads to:
\begin{align} 
\begin{split}
\mathrm{CCF}(\Delta t) &\approx \left( f_l(t+\Delta t) - f_k(t+\Delta t)\right)*f_k(t) \\ & \approx \mathrm{CCF}_{lk}(\Delta t)-\mathrm{ACF}_k(\Delta t).
\end{split}
\end{align}
This approximation was also used by \citet{Chelouche2012}. \newline
One complication to consider when calculating the CCF and ACF is the non-uniform time sampling being generic for astronomical data. In order to overcome this difficulty, some authors use the interpolated cross-correlation function method (ICCF, \citealt{Gaskell1987}), where mean and standard deviation of the time series are calculated at every time step, taking into account only the values within the overlapping apart of the light curves. Another complication arises from propagating the magnitude errors to errors for the time delay. Most of the CCF-related approaches have problems with doing so.\newline
Both problems can be solved with advanced reverberation mapping techniques based on fitting and modeling the light curves using a structure function model. What is explained here for the CCF and ACF, will also apply basically to more advanced reverberation mapping techniques.

%% file: section3.tex
\section{QUASAR LIGHT CURVES AS A STOCHASTIC PROCESS}
\label{sec:QuasarLightCurvesAsAStochasticProcess}

Simple interpolation methods fail when trying to carry out reverberation mapping on sparsely and non-uniform sampled measurements. Although there are some methods that can deal with some amount of non-uniformness, like the ICCF \citep{Gaskell1987}, we had shown that they are not suitable for our purposes. In Appendix \ref{sec:ResultsFromTestData} we demonstrate that these methods are not suitable for broadband data using tests of simulated data with non-uniform
time sampling. We need a description of the quasar variability that allows for reasonable interpolation on arbitrary times in between measurements. Following \citet{Kozlowski2009} and \citet{Butler2011}, we build a model for the quasar light curves based on a Gaussian process, because the Gaussian is the simplest two-point distribution function with a non-trivial variance that allows to fit and stochastically interpolate light curves.

Quasar light curves vary stochastically across a large dynamic range of time scales (e.g. \citealt{Kozlowski2009}). Their variability is sensibly characterized by a \textit{structure function} (e.g. \citealt{Hughes1992}, \citealt{Collier2001}, \citealt{Kozlowski2009}), which describes the mean squared difference (or, sometimes, root mean square difference) between pairs of observations of some object's brightness as a function of the time lag difference between the observations. In more detail, the structure function is a description of a second-order statistic of the brightness history of the source. As such, it does not give a direct description on how to fit such measurements or generate mock data.

A model and an algorithm based on this is built to have a consistent description of quasar variability, from which we can not only estimate structure function parameters of given light curves, but also generate mock light curves consistent with any reasonable set of structure function parameters, fit light curves and, as a main goal, produce a reverberation mapping model that is able to deal with very uneven time sampling as it is present in SDSS S82 quasar light curves. Because the Gaussian is the simplest two-point distribution function with a non-trivial variance that meet this conditions, we build this model from a Gaussian process.
The description here is mainly based on \cite{Butler2011} and notes by Bovy et al. (2011).

Assume a set of $N$ measurements $m_i$ taken at time $t_i$, being calibrated magnitude or flux measurements taken in a single bandpass of a single source associated with an uncertainty variance $\sigma_i$. The structure function $\mathrm{V}(|\Delta t|)$ is then defined \citep{Rybicki1992} as the expectation value $\mathrm{E}[\cdot]$ for the difference between observation $m_i$ and $m_j$ (with $i \neq j$),
\begin{equation}
 \mathrm{E}[(m_i - m_j)^2]=\sigma_i^2+\sigma_j^2 + \mathrm{V}(|t_i-t_j|).
\end{equation}
Here, the observations are presumed to be independent, and the structure function $\mathrm{V}(\cdot)$ effectively describes the variance.

To proceed, one must specify a concrete form for the quasar structure function, and two forms have been used in literature (e.g. \citealt{Schmidt2010}, \citealt{Butler2011}, \citealt{MacLeod2010}), first a power-law
\begin{equation}
\mathrm{V}(|\Delta t_{ij}|)=A^2 \left( \frac{\Delta t_{ij}}{1 \; \mathrm{yr}}        \right)^{\gamma}
\end{equation}
(e.g. \citealt{Schmidt2010}), where the amplitude $A$ quantifies the root-mean-square magnitude difference on a one year timescale, and $\gamma$ characerizes the time dependence of this difference. As $\mathrm{V}_{\Delta t= \infty} \rightarrow \infty$, for $\mathrm{V}_{\infty}$, it is practical to use reference values, $\mathrm{V}(t_{\mathrm{obs}})$ and $t_{\mathrm{obs}}$ in the characterization.

Second, one can describe quasar structure function as a damped random walk (DRW), for which the covariance function of a Gaussian process has an exponential form
\begin{equation}
 C_{ij} = \frac{\omega^2}{2} \exp \left( - \frac{|\Delta t_{ij}|}{\tau}    \right),
\end{equation}
(e.g. \citealt{Butler2011}) where $\tau$ is a damping time scale and $\omega^2$ is the intrinsic variance of the process.
Following \cite{MacLeod2012}, using the asymptotic value of the structure function $\mathrm{V}$ ($\mathrm{V}_{\infty}=\sqrt{2}\omega$) results in
\begin{equation}
\mathrm{V}(|\Delta t_{ij}|)=  \frac{ \omega^2}{2} \left( 1 - \exp \left(   \frac{-2|\Delta t_{ij}|}{\tau} \right)   \right).
\end{equation} 
The DRW model can be equivalently parameterized $\tau$ and the slope of $\mathrm{V}$ on short time scales, $\hat{\omega}=\sqrt{2\omega^2 / \tau}$ \citep{Kelly2009}. \\
A detailed description can be found in Appendix \ref{sec:TheStructureFunctionAndTheGaussianProcess}.

Depending on application, $|\Delta t_{ij}|$ can refer to the time lag between observations in the quasar rest frame or in the observed frame. Referring to the quasar rest frame, what needs a priori knowledge of the quasar redshift, can be important if the structure function parameters being estimated should be linked to physical properties of a quasar.

There is some discussion on which structure function model would fit best. \cite{Kelly2009}, \cite{Kozlowski2009}, \cite{MacLeod2012} and \cite{Andrae2013} have shown that quasar variability is well modeled by the DRW. \cite{Zu2013b} tested whether the DRW model provides an adequate description of quasar variability across all time scales. On time scales larger than a few years, the light curves are generally consistent with the DRW model but are not giving clear constraint on models. Alternatively, some authors \citep{Hook1994, Richards2006, Richards2008, Schmidt2010, Morganson2014} use the power-law model described above. \\

The structure function is the basis for the Gaussian process model that we fit to the data.

A Gaussian process is characterized by a function describing the mean measurement $\mathbf{m}(t)$ (magnitude or flux) as a function of time $t$ and a function $C(t,t')$ describing the covariance between observations at different epochs $t$ and $t'$. Assuming the mean is constant and the process is stationary such that $C(t, t') \equiv C (t-t')$, the probability of a set of $N$ observations $\left\{ m_i\right \} ^N _{i=1}$ is given by that of the $N$-dimensional Gaussian with mean $(m, m, ..., m)^{\mathrm{T}}$ and $N \times N$ dimensional covariance matrix $C$ with elements $C_{ij}=C(t_i - t_j)$.

After parameterizing the structure function, the complete model - the Gaussian process with mean vector $\bar{\mathbf{m}}$  and variance V - for any set of observations is specified by only three model parameters, either in the case of the power law $(\bar{m},A,\gamma)$ or in the case of a damped random walk $(\bar{m}, \omega, \tau)$. Thus, in turn, the likelihood $P(data|modelpar)$ can be described as $P(\mathbf{m}, A, \gamma) = \mathscr{N}(\mathbf{m}|\bar{m},C)$ or  $P(\mathbf{m}, \omega, \tau) = \mathscr{N}(\mathbf{m}|\bar{m},C)$, respectively, with $\mathrm{V}$ here expressed as a function of the structure function parameters $(A, \gamma)$ or $(\omega, \tau)$, respectively. The term $\mathscr{N}(\mathbf{m}|\bar{m},C)$ is the Gaussian process. This approach can yield posterior probability distribution to the two model parameters, $A$ and $\gamma$ or $\omega$ and $\tau$. We assign uninformative priors for the parameters, and then explore the posterior distribution for these parameters via a Markov Chain Monte Carlo (MCMC) approach. With using this, one can (a) model the light curve to get an estimate for the structure function parameters, (b) use the estimates for the structure function parameters e.g. for selecting quasars, or for advanced reverberation mapping techniques as in the present work that require interpolation of the light curves, (c) generate mock light curves as test data for this methods.

In the following, we refer to the formalism by \citealt{Rybicki1992} and \citealt{Zu2011} and summarize them here for convenience.

In practice, we marginalize over the mean $\bar{m}$
rather than fitting for it. This marginalization can be done analytically
when assuming a uniform prior on the mean (see \citealt{Rasmussen2006}, Equ. (2.45)) and leads to
the probability
\begin{align} \label{eqn:parameterlikelihoodquote1}
\begin{split}
P(\mathbf{m}|\mathbf{p}) & \propto \mathcal{L}(\mathbf{m}|\mathbf{p}) \\ & \equiv |S+N|^{-1/2} |L^{\mathrm{T}} C^{-1} L|^{-1/2} \exp \left( - \frac{\mathbf{m}^{\mathrm{T}} C_{\perp}^{-1}  \mathbf{m}}{2}  \right),
\end{split}
\end{align} 
where for the damped random walk model the remaining parameters $\mathbf{p}$ are $\tau$ and $\omega$ and for the power-law model $A$ and $\gamma$. $\mathcal{L}$ represents the likelihood function we are to maximize in order to find the most likely combination of those parameters. 

In Equ. \eqref{eqn:parameterlikelihoodquote1}, the intrinsic variability has a covariance matrix $S= \langle \mathbf{s}\mathbf{s} \rangle$, whereas the noise has a covariance matrix $N= \langle \mathbf{n}\mathbf{n}\rangle$. The covariance function of the  Gaussian process is then given by $C= S+N$. The component of the covariance matrix $C$ that is orthogonal to the fitted linear functions is given by $C_{\perp}^{-1} \equiv C^{-1} - C^{-1}L C_q L^{\mathrm{T}} C^{-1}$. $\mathbf{m}$ is the data vector. $L$ is a response matrix (see \cite{Zu2011} and \cite{Press1992}).

Suppose we have measured data $\mathbf{m}$ consisting of an underlying true signal $\mathbf{s}$, measurement uncertainties  $\mathbf{n}$ and a general trend defined by the response matrix $L$ and a set of linear coefficients  $\mathbf{q}$, thus,  $\mathbf{m}= \mathbf{s}+\mathbf{n}+L\mathbf{q}$. Using the linear coefficients to optimally determine the light curve mean, in the case of one light curve, we have one linear coefficient $q_1 \equiv \mathbf{q}$ for the mean, and the response matrix is simply a column vector $L_{i1}=1$ with an entry for each of the $K$ data points, $i=1,\cdots,K$. If we have two light curves with a possible offset in their means, we could use separate means for each of them, $(L_{i1},L_{i2})=(1,0)$ for data from the first light curve and $(L_{i1},L_{i2})=(0,1)$ for the second one. Additionally, $L$ can be used for light curve de-trending. For details on how we implemented de-trending, see Section \ref{subsub:EstimatingTheTimeDelay}.

The Gaussian-process formalism also allows
straightforward interpolation of the observed light curve between time
samples, with interpolation uncertainties, or the construction of mock
light curves with a given structure function. We use the latter below to
generate mock light curves to test our photometric reverberation mapping
technique. This formalism is explained in detail in \citet{Rasmussen2006} and 
\citet{Rybicki1992}. We refer the reader to those references for full
details.

\begin{figure*} 
\centering
\subfigure{\includegraphics{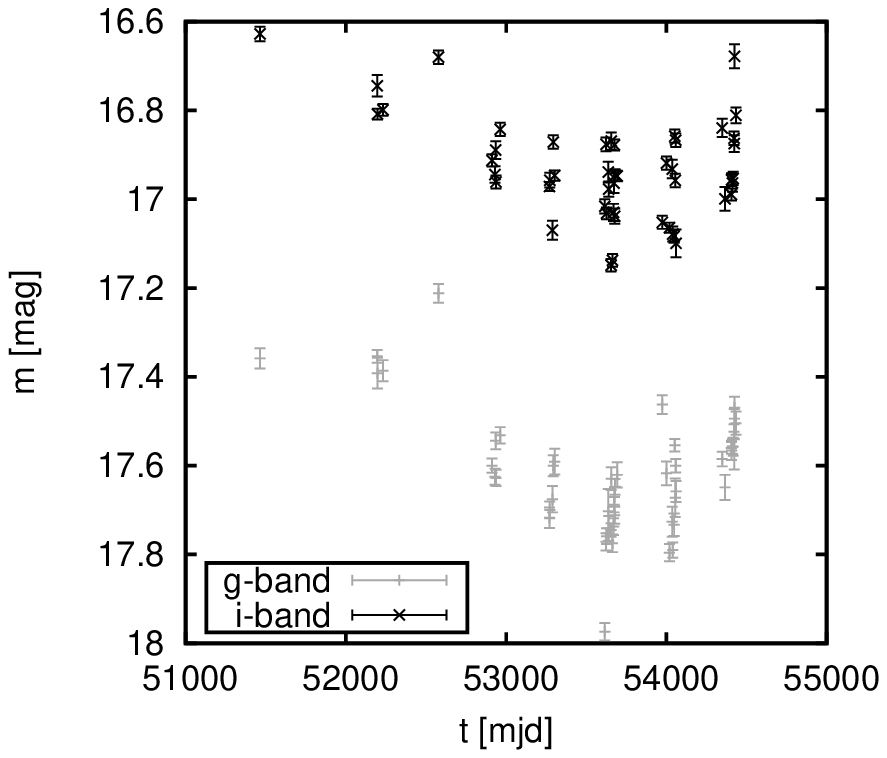}}\\
\subfigure{\includegraphics{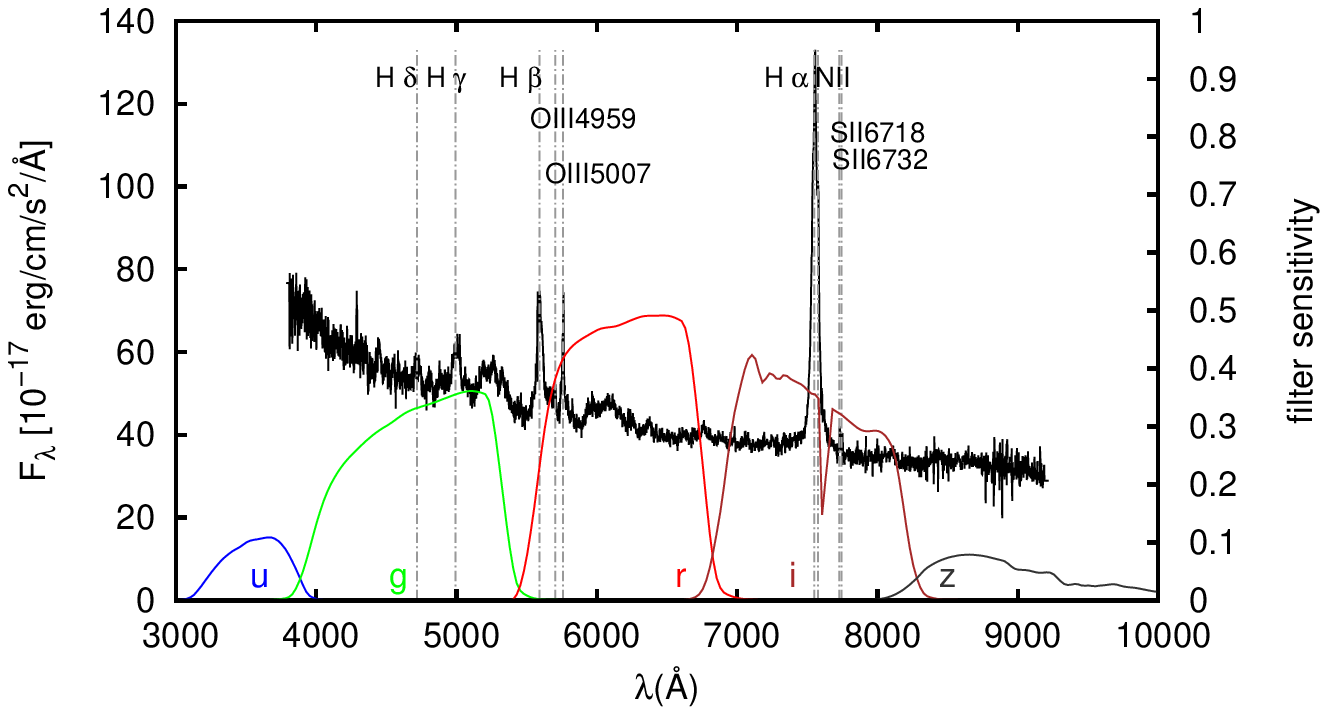}} 
\caption{Magnitudes in two filter bands of the example light curve \mbox{headobjid=587731185126146081} and corresponding spectrum plate=383, fiber=257, mjd=51818; the light curve is from a spectroscopically confirmed quasars of the SDSS Stripe 82 in a redshift region where \textit{g} band reflects almost exclusively accretion disk continuum emission, whereas the \textit{i} band has H$\alpha$ emission line contribution. Additionally, other emission lines are present.}
\label{fig:lc_587731185126146081}
\end{figure*}

\begin{figure*} 
\centering
\subfigure[]{\includegraphics{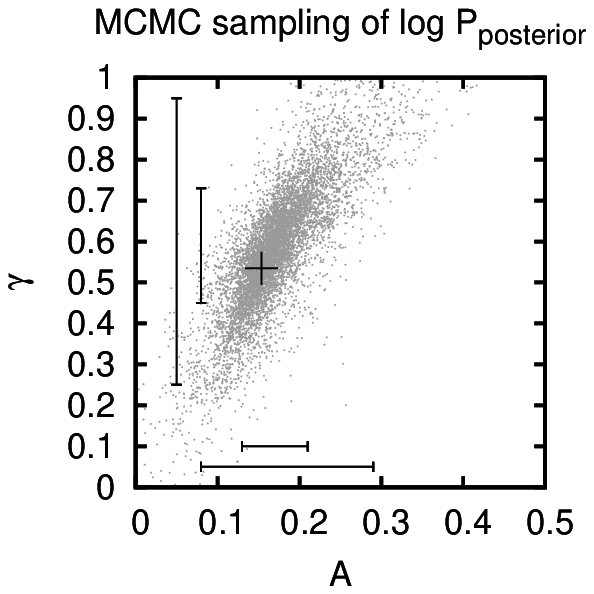}    }
\subfigure[]{\includegraphics{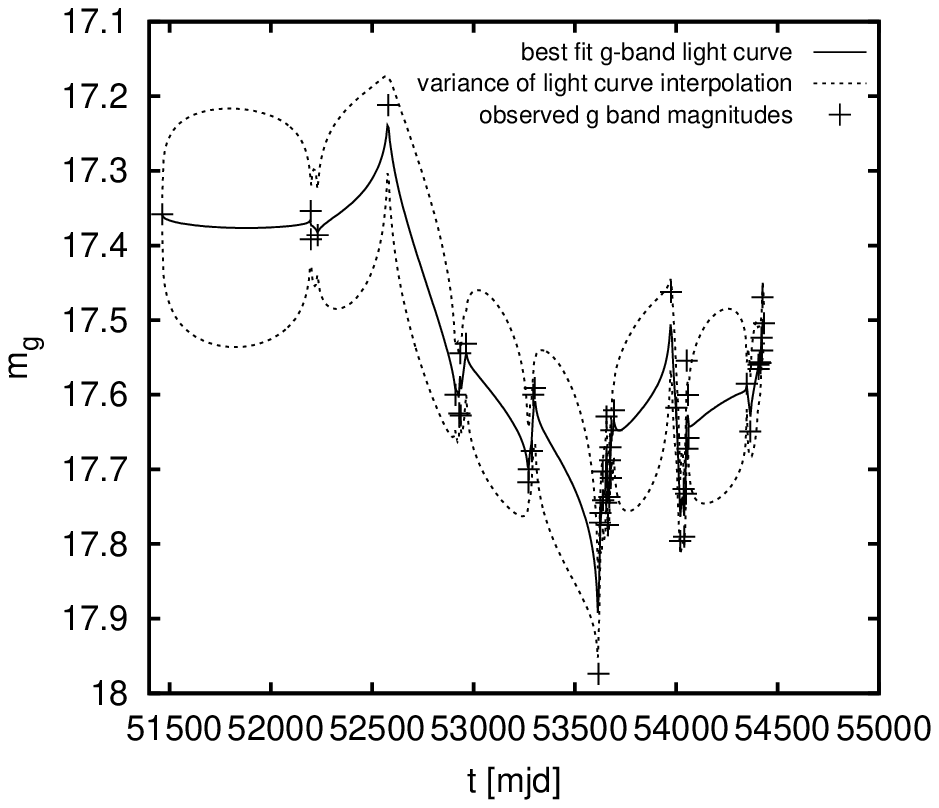}    }
                \caption{a) Result of a structure function model fit to the \textit{g} band light curve of the Stripe 82 quasar headobjid=587731185126146081, Fig. \ref{fig:lc_587731185126146081}; for this object the \textit{g} band
reflects almost exclusively accretion disk emission. Shown is the posterior probability distribution (PDF) obtained through MCMC \citep{Foreman2012}. The marginalized 68\% and 90\% confidence intervals for $A$ and $\gamma$ are indicated by horizontal and vertical bars. The cross marks maximum-at-posterior. \newline
b) Result of the interpolation of continuum light curve (\textit{g} band) for fit (Equation \eqref{eqn:bestestimatemeanlc}) for object headobjid=587731185126146081, 
Fig. \ref{fig:lc_587731185126146081}, derived from the best fit (structure function parameters at maximum at posterior) to the light curve's structure function. The solid line represents the best fit mean model light curves from the power law model.
The area between the dashed lines reflects the variance for the light curve prediction, arising from the stochastic models; this variance reduces to the range of
measurement errors at epochs where data exist (see Equation \eqref{eqn:bestestimatemeanlc}).\newline
These Figures can be also found as part of Figure \ref{fig:flow_noheader} in Section \ref{sec:StochasticReverberationMapping}, giving an overview about methodology and output of the stochastic reverberation mapping algorithm.}
\label{fig:plotlogl_modelpar_fitcontinuum_lc_587731185126146081}
\end{figure*}

For an example light curve (Fig. \ref{fig:lc_587731185126146081}), this fit is shown in the left panel of Fig. \ref{fig:plotlogl_modelpar_fitcontinuum_lc_587731185126146081}. We also give the best model parameter values along with the confidence regions (see below) in the right panel of Fig. \ref{fig:plotlogl_modelpar_fitcontinuum_lc_587731185126146081}.

The expression in terms of a Gaussian process allows one to generate a wide variety of mock light curves as test data for applications dealing with light curves, such as structure function parameter estimation or reverberation mapping.

As the fundamental property of a Gaussian process is that all of its marginal distributions - marginalizing over unobserved times - are Gaussian, generating a mock light curve is then just sampling from the appropriate Gaussian distribution. Realistic values in the power law case are 0.07 $< A <$ 0.28, 0.15 $< \gamma <$ 0.5 \citep{Schmidt2010}. In power law model, the amplitude $A$ quantifies the root-mean-square magnitude difference on a one year timescale. $\gamma$ is the logarithmic gradient of this mean change in magnitude. In DRW model, a larger $\omega$ makes curve more variable, a larger $\tau$ makes it more smooth (variability on longer timescales). Realistic values in the DRW case are 0.1 $< \omega <$ 0.4, 1 $< \log \tau < $ 3 \citep{MacLeod2010}.

%% file: section4.tex
\section{STOCHASTIC REVERBERATION MAPPING}
\label{sec:StochasticReverberationMapping}

As reverberation mapping has often carried out using CCF and ACF, a complication to consider when calculating the CCF and ACF is the non-uniform time sampling being generic for astronomical data. Also, in order to estimate the time delay and its uncertainty, we need to propagate the magnitude errors in the light curves to errors for the time delay. Most of the CCF-related approaches have problems with doing so, as they are not able to propagate errors. Additionally, in Appendix \ref{sec:ResultsFromTestData} we demonstrate that for a S82-like time sampling, these approaches are
not sufficient. Both problems can be solved by advanced reverberation mapping techniques based on fitting and modeling the light curves using a structure function model so the structure function parameters, the time lag $\tau_{\mathrm{delay}}$, its statistical confidence limits and in some cases additionally values are estimated.

In detail, we follow the approach of \cite{Rybicki1994} and \cite{Zu2011} that we extended for application to broad band photometry. Here, the basic methodology is described. In more detail, it is outlined in Appendix \ref{sec:TheBasicStochasticApproach} where the mathematical framework and the application of the method to data are described. Additionally, the methodology and output of the stochastic reverberation mapping algorithm is summarized in Figure \ref{fig:flow_noheader}.

\begin{samepage}The approach being described in this section is capable of
\begin{compactenum}[(i)]
\item handling transfer functions $\Psi(\tau_{\mathrm{delay}})$ instead of simply a $\tau_{\mathrm{delay}}$, thus being able to map out the physical structure of the broad line region that cannot be simply modeled by a $\delta$ function
\item not only interpolating between data points, but also making self-consistent estimates and including these uncertainties in the interpolation,
\item separating light curve means and systematic errors in flux calibration from variability signals and measurement uncertainties in a self-consistent way,
\item deriving simultaneously the lags of multiple emission lines and their covariances,
\item providing statistical confidence limits on all estimated parameters.
\end{compactenum}
\end{samepage}
The approach assumes that all emission-line light curves are scaled, smoothed and displaced versions of the continuum. We assume for simplicity that we have photometric quasar light curves in the \textit{k} (e.g., SDSS \textit{r}) and \textit{l} (e.g., SDSS \textit{g}) bands, where the \textit{l} band contains emission line and continuum flux, while the \textit{k} band has continuum only. Then we can write the fluxes as

 \begin{equation}
    \begin{split}
 f_k(t) & =  f_k^c(t) \\
 f_l(t) &=   f_l^c(t) +  f_l^e(t) \\
  &= s \, f_k^c(t) + e \int \Psi(\tau_{\mathrm{delay}})f_k^c(t-\tau_{\mathrm{delay}}) \,  \mathrm{d}\tau_{\mathrm{delay}} .
    \end{split}
 \label{eqn:fxfymodela}
  \end{equation} 
This equation is the general version of \eqref{eqn:rgmodel}, allowing allowing for a smoothed response due to arbitrary transfer functions.
In \eqref{eqn:fxfymodela}, $f_k$ and $f_l$ are the total fluxes in the \textit{k} and \textit{l} band respectively, and superscripts $^c$ and $^e$ denote continuum and emission-line contributions. $s$ and $e$ are linear scaling factors between \textit{k} and \textit{l} band variability. In our application, they are constrained spectroscopically (see Equ. \eqref{eqn:intfore} and \eqref{eqn:intfors}).
The delayed response to the continuum is described by the normalized one-dimensional transfer function $\Psi(\tau_{\mathrm{delay}})$ \citep{PetersonBook}, i.e., 
\begin{equation}
f^e(t)= \int_{-\infty}^{+\infty} \Psi(\tau_{\mathrm{delay}})f^c(t-\tau_{\mathrm{delay}}) \,  \mathrm{d}\tau_{\mathrm{delay}}.
\label{eqn:LAsIntegral}
\end{equation}

The generalization of this formalism to the case of two or more emission lines in separate bands is straightforward. We have not used this here, as it was not appropriate for the light curves we had analyzed. In the case of a $\delta$ function transfer function, Equ. \eqref{eqn:fxfymodela} reduces to Equ. \eqref{eqn:rgmodel}.

We assume that the quasar continuum light curve can be described as a Gaussian stochastic process (e.g., \citealt{Kozlowski2009}, \citealt{MacLeod2012}) and that the \textit{l} band flux varies linearly with the \textit{k} band flux \citep{Schmidt2012}. The continuum model is then characterized by a variance matrix $C_{kk}^{cc}$
resulting from any Gaussian stochastic variability process, e.g., the damped random walk \cite{Kelly2009} or a power-law structure function model \citep{Schmidt2010}. The emission-line covariance matrix $C_{ll}^{ee}$ is then given by
 \begin{equation}
    \begin{split}
C_{ll}^{ee}(\Delta t) &= \langle  f_l^e(t),f_l^e(t+\Delta t)    \rangle   \\
                      &= e^2 \int  \mathrm{d}\tau_{\mathrm{delay},1} \int  \mathrm{d}\tau_{\mathrm{delay},2}  \Psi(\tau_{\mathrm{delay},1})\Psi(\tau_{\mathrm{delay},2})\\ & \; \; \;  \; \;C_{kk}^{cc}(\Delta t - \tau_{\mathrm{delay},2}-\tau_{\mathrm{delay},1}).
   \end{split}
  \end{equation} 

Similarly, the continuum--line-emission cross terms are given by
\begin{align}
C_{ll}^{ec/ce}(\Delta t) &= e \int  \mathrm{d} \tau_{\mathrm{delay}} \Psi(\tau_{\mathrm{delay}})C_{ll}^{cc}(\Delta t \pm \tau_{\mathrm{delay}}) \nonumber \\
                         &= s^2 \, e \int  \mathrm{d}\tau_{\mathrm{delay}} \Psi(\tau_{\mathrm{delay}}) C_{kk}^{cc}(\Delta t \pm \tau_{\mathrm{delay}}) \\
C_{lk/kl}^{ec/ce}(\Delta t) &= e \int  \mathrm{d} \tau_{\mathrm{delay}} \Psi (\tau_{\mathrm{delay}}) C_{kk}^{cc}(\Delta t \pm \tau_{\mathrm{delay}})
\end{align}
where the $\pm$ refers to combinations in the sub- and superscripts of the left-hand side as $+/-$ and $C_{ll}^{cc} = s^2 C_{kk}^{cc}$, $C^{cc}_{kl} = s C^{cc}_{kk}$ as given by the flux model Equ.  \eqref{eqn:fxfymodela}.

Corresponding equations where the integrals are written-out using a $\delta$-function transfer function and the power law model can be found in the Appendix at \eqref{eqn:covxxfluxpowerlaw} \eqref{eqn:covxyfluxpowerlaw} \eqref{eqn:covyyfluxpowerlaw}. Also, Figure \ref{fig:flow_noheader} gives an overview on the usage of the different covariance matrices.

These terms can now be used to write down the covariance matrix for the \textit{k} band continuum and \textit{l} band continuum plus emission line fluxes as
\begin{equation}
C = \begin{pmatrix} C_{kk}^{cc} &  C_{kl}^{c,(e+c)} \\  \; \; \;  \; \,  C_{lk}^{(e+c),c} &  \; \; \; \; \, C_{ll}^{(e+c),(e+c)}  \end{pmatrix} .
\label{eqn:covariancematrixforlag}
\end{equation} 
with
\begin{equation}
C^{c,(e+c)}_{kl} = C^{ce}_{kl} + C^{cc}_{kl}
\label{eqn:cfxfy}
\end{equation}

\begin{equation}
 C^{(e+c),(e+c)}_{ll} = C^{cc}_{ll} + C^{ec}_{ll} +C^{ce}_{ll} + C^{ee}_{ll}.      
\label{eqn:cfyfy}  
\end{equation}

Using the covariance matrix as defined above, in Gaussian statistics the probability of some parameters (the structure function parameters and the time delay $\tau_{\mathrm{delay}}$) given the data (in flux units as we refer to flux here) can be computed, which yields a maximum likelihood approach $P(\mathbf{m}|\mathbf{p}) \propto \mathcal{L}(\mathbf{m}|\mathbf{p})$ (see Equ. \eqref{eqn:parameterlikelihood}) where $\mathbf{p}$ are the model parameters, i.e. the structure function parameters and $(e,s,\tau_{\mathrm{delay}})$ where $e$ and $s$ are constrained spectroscopically. How this approach is carried out technically, is shown in Appendix \ref{sec:TheStructureFunctionAndTheGaussianProcess} and \ref{sec:ReverberationMappingCovarianceMatrixElements}.

\begin{figure*}
\centering
              \includegraphics[trim=1.1inch 1.5inch 1.5inch 1.2inch, clip=true]{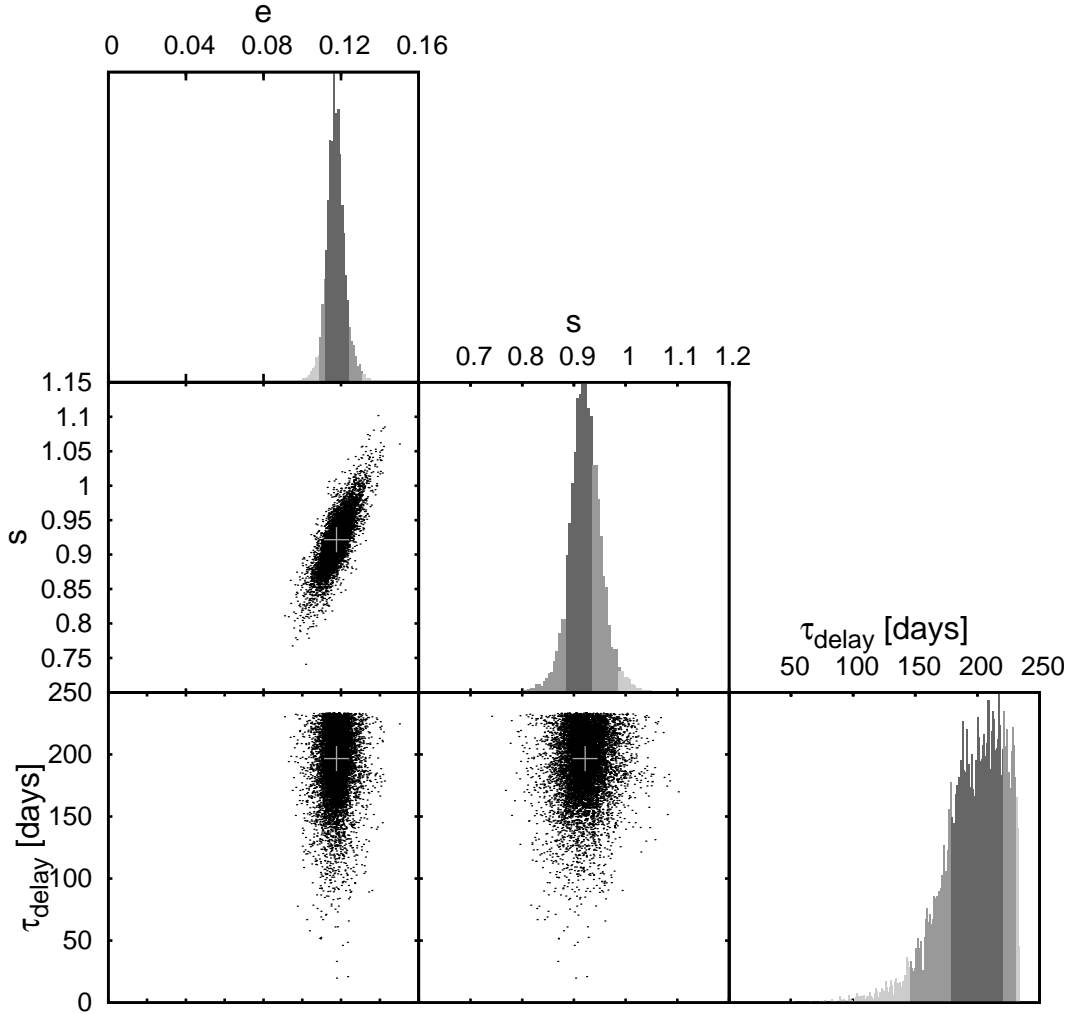}    
                \caption[example light curve]{\footnotesize{PDFs, represented in a 'triangle plot' for the estimate of the reverberation delay $\tau_{\mathrm{delay}}$ for the quasar from Fig. \ref{fig:lc_587731185126146081}, derived from the simultaneous fit to the light curve that is in one band only continuum (\textit{g} band), and in the other band (\textit{i} band) a combination of continuum and delayed emission line flux (see Equ. \eqref{eqn:lagposterior} in Appendix B). The figure shows a MCMC sampling of the PDF for the spectroscopically constrained emission-line fraction $e$ and the emission line delay $\tau_{\mathrm{delay}}$, along with their marginalized 68\% and 95\% confidence regions (grey shading); cross marks maximum-at-posterior $\tau_{\mathrm{delay,MAP}}$=195.084, $e_{\mathrm{MAP}}$=0.1176 }    }
   \label{fig:plotlogl_lag_elc_587731185126146081}
\end{figure*}

To illustrate the typical shape of the probability distribution functions (PDFs), an example output is shown in Fig. \ref{fig:plotlogl_lag_elc_587731185126146081}. 

\subsection{Parameter Estimation by MCMC}

We "chain" the likelihoods, as we first analyze the continuum light curve on its own to estimate the structure function parameters $\mathbf{p}_{\mathrm{struc}}$. Then we do a joint analysis of the continuum and the emission line light curve using the values for the structure function parameters estimated in the first step in order to estimate $\mathbf{p}_{\mathrm{trans}}$.

\subsubsection{Estimating Structure Function Parameters}

Estimation of the structure function parameter is done by evaluating the logarithmic posterior probability distribution
\begin{equation}
\log P_{\mathrm{posterior}} = \log P(\mathbf{p}) + \log \mathcal{L}(\mathbf{m}|\mathbf{p}) 
\label{eqn:logposteriorstrucfunc}
\end{equation}
where $\mathbf{p}=\mathbf{p}_{\mathrm{struc}}$ are the structure function parameters and $\mathbf{m}$ the measured light curve points. 
$\mathcal{L}(\mathbf{m}|\mathbf{p})$ is given by Equ. \eqref{eqn:parameterlikelihood} and $P(\mathbf{p})$ represents the prior PDF of the structure function parameters. In this equation, the intrinsic variability of the data is described by a covariance matrix $S$, whereas the noise has a covariance matrix $N$. For estimation of the structure function parameters, we use a data vector 
$\mathbf{m} \leftarrow \mathbf{f_k}$.

For a power law model, we have 
\begin{equation}
\log P(\mathbf{p}) = \log P(A) + \log P(\gamma),
\label{eqn:powerlawmodelposterior}
\end{equation}
where
\begin{align}
 P(A) &\propto \left\{\begin{array}{cl} \frac{1}{A}, & \mbox{if } 0 < A \leq 1     \\ 0, & \mbox{else} \end{array}\right. \\
 P(\gamma) &\propto \left\{\begin{array}{cl} \frac{1}{1+\gamma^2}, & \mbox{if } 0 \leq \gamma \leq 1     \\ 0, & \mbox{else} \end{array}\right.  .
\end{align}
In this way we enforce our assumption that the power law exponent $\gamma$, the logarithmic gradient of this mean change in magnitude, is positive and that the average variability on a 1 year timescale is less than 1 magnitude.

Based on the tests described in Appendix \ref{sec:TestData}, we use the Affine Invariant Markov chain Monte Carlo (MCMC) Ensemble sampler \citep{Foreman2012} to explore the posterior probability distribution.

As a result of testing, for the estimation of the structure function parameters and the mean magnitude $\bar{m}$, we use 10 walker\footnote{\label{foot:foreman} A detailed description of this terms and the algorithm can be found in \citep{Foreman2012}.}, up to 1000 iterations in a burn-in\footnotemark[\value{footnote}] run, 200 iterations in a post-burn-in\footnotemark[\value{footnote}] run, initialization $\mathbf{x}^{(0)}=(A^{(0)}, \gamma^{(0)}) = (0.1,0.1)$. As $\bar{m}$ is an output parameter of the algorithm evaluating the likelihood function, it is not a component of the vector $\mathbf{x}$.

\subsubsection{Estimating the Time Delay}
\label{subsub:EstimatingTheTimeDelay}

Estimation of the time delay is done by evaluating the logarithmic posterior probability distribution
\begin{align} \label{eqn:lagposterior}
\begin{split}
\log P_{\mathrm{posterior}} &= \log P(\mathbf{p}_{\mathrm{trans}})+ \log(\mathcal{L}(\mathbf{m}|\mathbf{p}_{\mathrm{trans}})) \\
 &= \log P(\tau_{\mathrm{delay}},e,s) \\ & \; \;\; + \log \mathcal{L}(\mathbf{m}_x, \mathbf{m}_y|\tau_{\mathrm{delay}},e,s).
\end{split}
\end{align} 

$\mathcal{L}(\mathbf{m}|\mathbf{p}_{\mathrm{trans}})$ is given by \eqref{eqn:parameterlikelihood} and $P(\tau_{\mathrm{delay}},e,s)$ represents the prior PDF of the transfer function parameters.

$P(\tau_{\mathrm{delay}},e,s)$ consists of a prior on the time delay, $P(\tau_{\mathrm{delay}})$, and a prior on $(e,s)$.\newline
The prior on the time delay is given by
\begin{align} \label{eqn:priorgauss}
\begin{split}
 P(\log_{10} \tau_{\mathrm{delay}}) \propto  \left\{\begin{array}{cl} & \frac{1}{\sqrt{2\pi}} \exp \left[ - \frac{(\log_{10}\tau_{\mathrm{delay}}-\log_{10}\tau_{\mathrm{delay,0}})^2}{\log_{10}2}  \right], \\& \mbox{if } 0.25\times\tau_{\mathrm{delay,0}} < \tau_{\mathrm{delay}} < 4\times\tau_{\mathrm{delay,0}}    \\ &0,  \mbox{else} \end{array}\right. .
\end{split}
\end{align} 

$\tau_{\mathrm{delay,0}}$ is set to the time delay inferred from the Kaspi relation Equ. \eqref{eqn:kaspi_r_blr_5100} if $\lambda L_{\lambda}(5100\; \mathrm{\mathring{A}})$ is available, and from the virial assumption otherwise.

For the virial mass estimates, it has been assumed \citep{Vestergaard2006} that the broad line region (BLR) is virialized, the continuum luminosity is used as a proxy for the BLR radius, and the broad line width (FWHM or line dispersion) is used as a proxy for the virial velocity. The virial mass estimate is then expressed as
\begin{align} \label{eqn:virialassumption}
\begin{split}
\log \left( \frac{M_{\mathrm{BH,vir}}}{\mathrm{M_{\sun}}}   \right) = & a+b \log \left( \frac{\lambda L_{\lambda}}{10^{44} \mathrm{erg \; s^{-1}}}  \right)   \\&+ 2 \log \left( \frac{\mathrm{FWHM}}{\mathrm{km \; s^{-1}}}     \right),
\end{split}
\end{align} 
where the coefficients $a$ and $b$ are empirically calibrated against local AGNs with RM masses or internally among different lines. This results in an expected rest-frame delay of
\begin{equation}
\tau_{\mathrm{delay,0}}=(1+z)  \cdot 10^{\mathrm{LOGBH}} \cdot \frac{5.121039 \; \mathrm{light days}}{\mathrm{FWHM\_BROAD\_HB}^2}.
\label{eqn:tau_delay_vestergaard}
\end{equation}
LOGBH and FWHM\_BROAD\_HB are the logarithmic virial $M_{\mathrm{BH}}$ and FWHM of broad H$\beta$ (km/s), respectively, from the Catalog of Quasar Properties from SDSS DR7 \citep{Shen2011}.

How $(e,s)$ can be constrained by a prior, depends much on the information that is available beside the photometric data. In the case of the SDSS S82 data, spectroscopic data are used to constrain $(e,s)$.

As we assume the flux model \eqref{eqn:fxfymodela}, we need to know which part of flux in the \textit{l} band belongs to the continuum and which is emission line contribution. From spectrum, we can get some information on $e$ and $s$, as
\begin{align}
e & \approx \frac{1}{f_k} \int_{l} f_e(\lambda) \omega _l (\lambda) \ \mathrm{d} \lambda  \label{eqn:intfore} \\
s & = \frac{1}{f_k}  \int_{l} f_c(\lambda) \omega _l (\lambda) \ \mathrm{d} \lambda . \label{eqn:intfors}
\end{align}
with
\begin{equation}
f_k = \int_{k } f_k (\lambda) \mathrm{d} \lambda
\end{equation}
and
$\omega _{l}(\lambda)$: filter curve in the $l$ band, normalized so $\int \omega _{l}(\lambda) \mathrm{d} \lambda=1$, $\mathrm{EW} = \int{ \frac{f_l^e(\lambda) - f_l^c(\lambda)}{f_l^c(\lambda)} } \mathrm{d}\lambda$: equivalent width of the line.

A fit for the emission line and the continuum has to be done to get $f_l^e$ and the continuum, $f_l^c \approx f_l - f_l^e$.
The emission line is fitted as a Gaussian, using the provided $\mathrm{FWHM_{line}}$ and the continuum level at the emission line, $f_l^c(\lambda_{\mathrm{line,obs}})$, 
\begin{align}
\begin{split} 
         f_l^e(\lambda) &=  \left(f_l(\lambda_{\mathrm{line,obs}})-f_l^c(\lambda_{\mathrm{line,obs}})\right) \\ & \; \; \; \cdot \exp\left(- \frac{ ( \lambda- \lambda_{\mathrm{line,obs}})^2}{2\sigma^2} \right) 
\end{split}
\end{align} 
with
\begin{equation}
\sigma=2.35 \; \mathrm{FWHM_{line,obs}}. 
\end{equation}
For the continuum level, we use the approximation  
\begin{align}
\begin{split} 
f_{\mathrm{cont}}(\lambda_{\mathrm{line,obs}}) &= \frac{f_{\mathrm{left}}+f_{\mathrm{right}}}{2} \\
&= \frac{1}{2} \left( f\left(\lambda_{\mathrm{line,obs}} - \frac{\mathrm{EW_{obs}}}{2}\right) \right.\\
&  \; \; \; \left.+f\left(\lambda_{\mathrm{line,obs}}  \frac{\mathrm{EW_{obs}}}{2}\right) \right).
\end{split}
\end{align} 

Applying this to \eqref{eqn:intfore} and \eqref{eqn:intfors} gives initial values $(e_0,s_0)$. We are now able to predict an equivalent width from some $(e,s)$ and compare it to the observed one,  
\begin{align}
\mathrm{EW}_{\mathrm{pred}}(t)=\frac{e}{s} \frac{\int{f_{\mathrm{line}}(\lambda) (t-\tau_{\mathrm{delay}})\mathrm{d} \lambda   }}{\int{f_{\mathrm{line}}(\lambda) (t-\tau_{\mathrm{delay}})\omega(\lambda) \mathrm{d} \lambda}} 
\label{eqn:ewpred}
\end{align}
where
\begin{equation}
f_{\mathrm{line}}(\lambda)(t)=a(t) \cdot f_{\mathrm{line}}(\lambda)(t_0)
\end{equation}
with $t_0$ is the time the spectrum was taken. As we assume $e$ and $s$ being constant over time, 
\begin{equation}
a(t)=\frac{f(t)}{f(t_0)}=\frac{f(m_x(t))}{f_x(t_0)},
\end{equation}
interpolation of the light curve in the continuum only band $x$ has to be done.

We are now able to predict an equivalent width from some $(e,s)$ and compare it to the observed one at the time $t_0$ the spectrum was taken. From this, with $\mathrm{EW}_{\mathrm{pred}} \equiv \mathrm{EW}_{\mathrm{pred}}(t_0)$ one gets the likelihood term
\begin{equation}
\mathscr{L}_{\mathrm{spec}} \propto \frac{1}{\sqrt{2 \pi} \delta \mathrm{EW}_{\mathrm{obs}}}\exp \left( - \frac{(\mathrm{EW}_{\mathrm{obs}}-\mathrm{EW}_{\mathrm{pred}})^ 2}{2 \delta EW_{\mathrm{obs}}^ 2}   \right),
\label{eqn:likspek}
\end{equation}
where $\delta EW_{\mathrm{obs}}$ is measurement uncertainty in the equivalent width of the observed emission line. In our application to SDSS S82 data, observed equivalent widths $\mathrm{EW}_{\mathrm{obs}}$ are retrieved from the Catalog of Quasar Properties from SDSS DR7 \citep{Shen2011}.

Equation \eqref{eqn:likspek} is multiplied with the previous likelihood term \eqref{eqn:parameterlikelihoodquote1}
to describe the likelihood of the model parameter $\mathbf{p}$ given the data $\mathbf{m}$. In this equation, the intrinsic variability of the data is described by a covariance matrix $S$, whereas the measurement uncertainties have a covariance matrix $N$. For estimation of the time delay from one emission line, we use a data vector 
$\mathbf{m} \leftarrow (\mathbf{f_k},\mathbf{f_l} )$.

Light curve de-trending is applied through response matrix $L$. Basically, if we have two light curves with a possible offset in their means, we could use separate means for each of them, $(L_{i1},L_{i2})=(1,0)$ for the continuum light curve $f_k$ and $(L_{i1},L_{i2})=(0,1)$ for the light curve containing a continuum and emission line contribution $f_l$.

As mentioned in some papers referring to the basic approach of this algorithm, e.g. \cite{Zu2011}, the response matrix $L$ can also be used to describe and in this way remove a general trend in the light curve, what is called de-trending. De-trending has been shown to considerably improve reverberation mapping, as removing a general linear trend in the light curve so realizes better the limit of stationary light curves (e.g. \citealt{Welsh1999}).
For de-trending, after some tests we decided to use $(L_{i1},L_{i2})=(t_i,0)$ for $f_k$ and $(L_{i1},L_{i2})=(0,t_j)$ for $f_l$.

\begin{figure*}
              \includegraphics[width=1.0\textwidth]{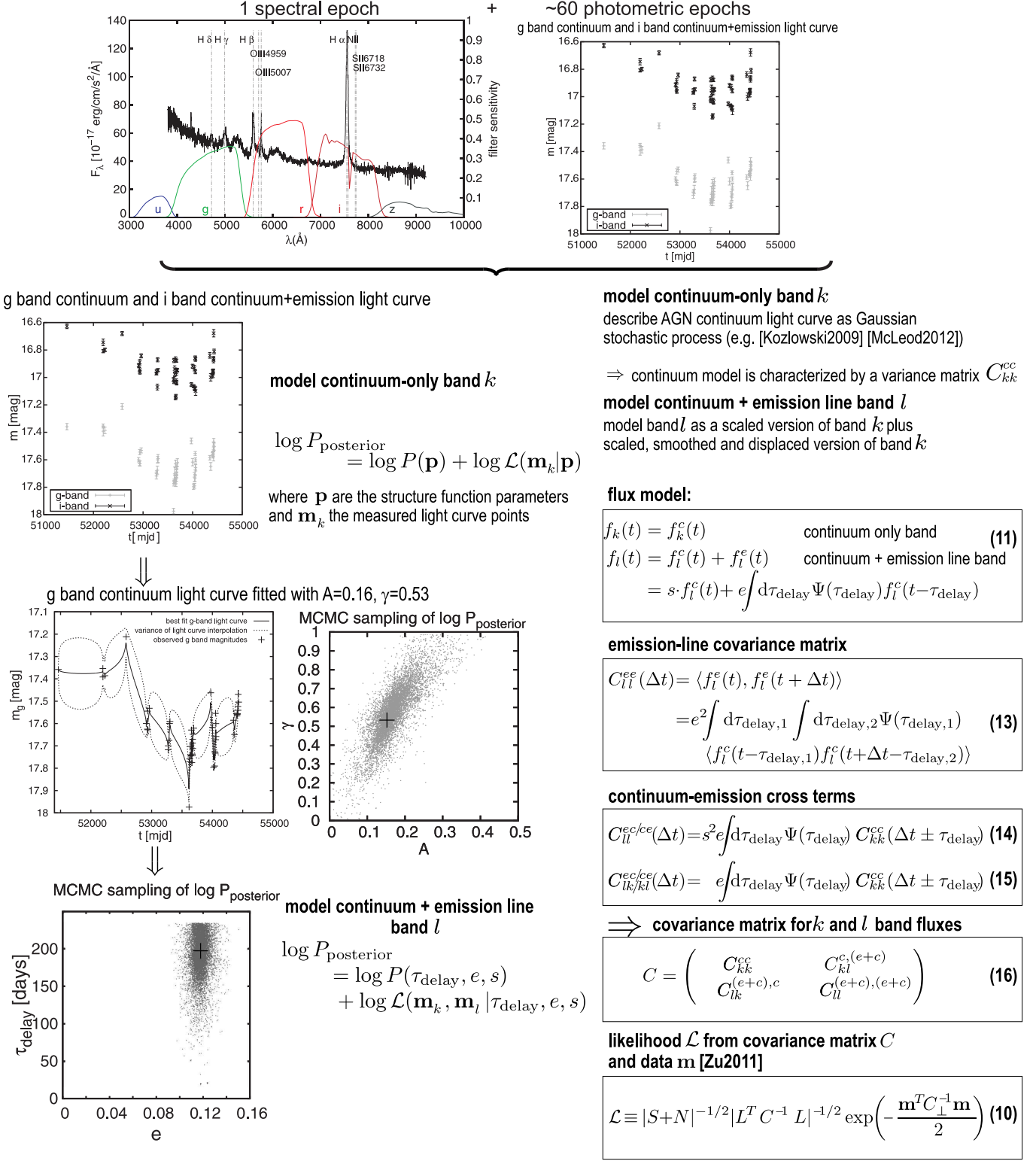}    
                \caption[Methodology of stochastic reverberation mapping
illustrated using the example of light curve headobjid=587731185126146081]{\footnotesize{Methodology of stochastic reverberation mapping
illustrated using the example of light curve headobjid=587731185126146081,
with z=0.1506, continuum-only band \textit{k}: \textit{g} band,
continuum+emission line band \textit{l}: \textit{i} band with strong H$\alpha$. 
}}
                \label{fig:flow_noheader}
\end{figure*}

Figure \ref{fig:flow_noheader} summarizes methodology and output of the stochastic reverberation mapping algorithm.

Based on the tests described in Appendix \ref{sec:TestData}, we use the Affine Invariant Markov chain Monte Carlo (MCMC) Ensemble sampler \citep{Foreman2012} to explore the posterior probability distribution.

For the estimation of $\tau_{\mathrm{delay}}$, $s$, $e$, we use 15 walker, up to 2000 iterations in a burn-in run, 800 iterations in a post-burn-in run,  initialization $\mathbf{x}^{(0)}=(s^{(0)}, e^{(0)}, \tau_{\mathrm{delay}}^{(0)})$, where $s^{(0)}$, $e^{(0)}$, $\tau_{\mathrm{delay}}^{(0)}$ depend on the current light curve.\newline
The start position of the walkers is $x^{(0)}+r$ where $r$ is some random number, so the walkers start in a small area in parameter space around $x^{(0)}$.

%% file: section5.tex
\section{QUASAR SPECTROSCOPY AND LIGHT CURVES IN THE SDSS STRIPE 82}
\label{sec:ApplicationToSDSSS82Data}
The Sloan Digital Sky Survey (SDSS) \citep{York2000} provides homogeneous and deep
($r <$ 22.5) photometry in five passbands (\textit{ugriz}) for typically over 60 epochs of observations over a decade in a 290 deg$^2$ area of the Southern Galactic cap known as Stripe 82 (S82), \citep{Frieman2008, Annis2011, Ivezic2012}. These photometry epochs were obtained in early "seasons" of about 2--3 months, effectively sampling time scales from days to years.

The photometric data for the \textit{ugriz} bands are simultaneous, and an example of such a light curve (only two bands are plotted) is shown in Fig. \ref{fig:lc_587731185126146081}. This area of SDSS is also exceptional in that it has complete spectroscopic quasar identification \citep{Shen2011}, resulting in a sample of 9156 quasars, with spectra. Additional information on theses quasars exists in the
Catalog of Quasar Properties from SDSS DR7 \citep{Shen2011}.

Until first data release of LSST, this data S82 data set, with its combination of single epoch spectra and multi-band light-curves for 10$^4$ quasars (see also Fig. \ref{fig:lc_587731185126146081}) is the best data set to carry out broad-band reverberation mapping. It is the same data set that has been used before by \cite{Schmidt2010}. 

The data are all publically accessible through the SDSS data archive \footnote{http://casjobs.sdss.org/CasJobs/default.aspx}. For our application, all light-curves have been cleaned of manifest outliers, by simply considering measurements having a magnitude error of $\geq$ 0.1 and remove them. To obtain further information on each individual quasar, such as for instance equivalent widths or values needed for calculating a prior on the size of the broad line region as described in Section \ref{subsub:EstimatingTheTimeDelay}, Eqn. \eqref{eqn:priorgauss}, we cross-matched the list of objects from SDSS Stripe 82 with the Catalog of Quasar Properties from SDSS DR7\footnote{the catalog is available at \newline \url{https://www.cfa.harvard.edu/~yshen/BH\_mass/data/catalogs/}} \citep{Shen2011}.

We linked the spectra, the light curves and the additional information from DR7 by requiring a positional match of $\sim$1". For 9120 light curves from SDSS Stripe 82, a unique corresponding entry was found in the DR7 Quasar Properties Catalog. For the remaining 36 light curves, no match was found. There were no double matches.

For the subsequent analysis, we convert all light curve measurements to linear fluxes rather than magnitudes. To get the flux $f_k$ in the specific band \textit{k} in units of erg s$^{-1}$ cm$^{-2}$, the effective bandpass width $w_{\lambda,k}$ has to be taken into account, resulting in
\begin{align}
\begin{split} 
f_k(t) &= 3631 \times 10^{-23}  \frac{2b_k c}{w_{\lambda,k}} \\& \;\; \;  \times \sinh \left( - \frac{ m_k(t)  \log(10)}{2.5} -\log(b_k)  \right)  \; \mathrm{erg \, s^{-1}cm^{-2}}.
\end{split}
\end{align}

Not all of the 9,156 quasars in the sample are comparably suitable for broad-band reverberation mapping. Mock data analyses (see Appendix \ref{sec:ResultsFromTestData}) have shown that two conditions need to be satisfied: a redshift interval where one band as important emission line contributions, while another band is free of them; sufficiently many photometric epochs; in S82 they range from 2 to 160, with a median of 66.

To identify suitable redshift ranges that maximize continuum-line contrast between two different bands, we consider the H$\alpha$, H$\beta$, \ion{Mg}{2} lines for reverberation mapping and the \ion{C}{4}, \ion{N}{2} $\lambda$6585, \ion{S}{2} $\lambda$6718, \ion{S}{2} $\lambda$6732, \ion{O}{3} $\lambda$4959, \ion{O}{3} $\lambda$5007, Ly$\alpha$, Ly$\beta$, \ion{C}{3}], \ion{Fe}{2} $\lambda$2382, \ion{Fe}{2} $\lambda$2600, H$\gamma$, H$\delta$ lines as contaminants, with the data taken from Table 2 in \cite{VandenBerk2001}. To check whether a line falls into a band or not, we define the limits of a band as the wavelength where the neighboring filters have transmission 0.
This results in relatively narrow redshift ranges having only one out of H$\alpha$, H$\beta$, \ion{Mg}{2} in one band (the continuum+emission line band) and at least one other band free of all of them. There redshift ranges are summarized in Table \ref{tab:linecontribution}.

Within these redshift ranges, we then identify quasars whose light-curve sampling is relatively good, as this has great influence on the possibility of estimating a certain time delay $\tau_{\mathrm{delay}}$. In evenly sampled data, the sampling rate must be higher than the (expected) time delay. In unevenly sampled data, there must be at least some time intervals smaller than or equal to the (expected) time delay.

\begin{figure}[!ht]
\begin{center}    
\subfigure[ ]{\includegraphics{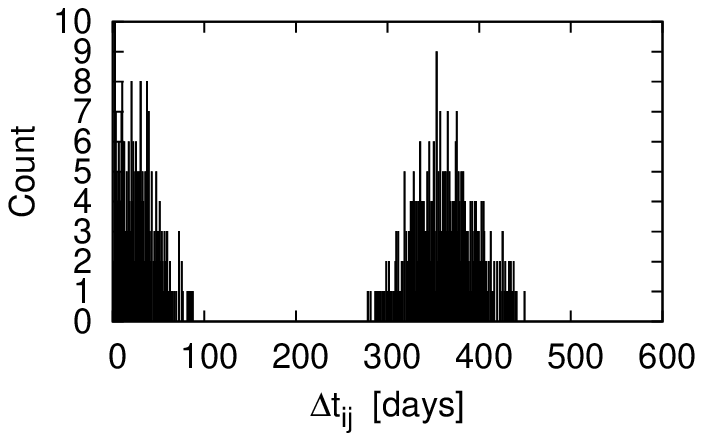}}\qquad
\subfigure[ ]{\includegraphics{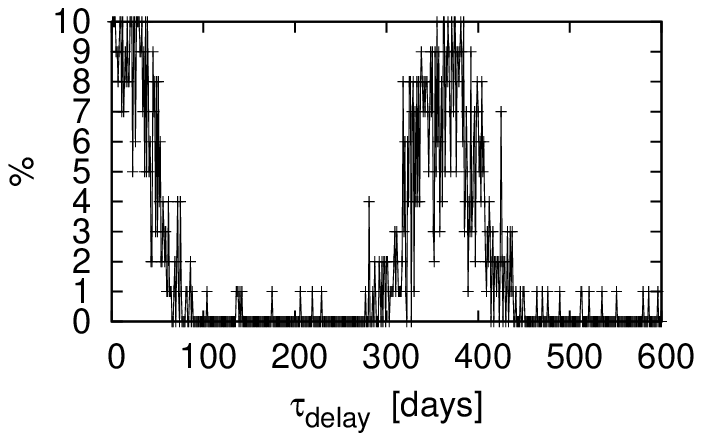}}
\caption[histogram of observational time lags]{\footnotesize{(a) Histogram of the observational time lags (first 600 days) in the light curve (differences between the data points); total observational period: 2959 days,\newline(b) percentage recovered $\pm$ 1 day vs. $\tau_{\mathrm{delay}}$ from mock light curves with the time sampling resulting to observational time lags from figure (a); crosses mark the evaluated lags, lines are displayed to guide the eye.
}}
\label{fig:percrec}
 \end{center}
\end{figure}

We illustrate this in Fig. \ref{fig:percrec}, by generating mock light curves within the actual S82 time sampling, but varying $\tau_{\mathrm{delay}}$. By applying the likelihood approach described in Section \ref{sec:StochasticReverberationMapping}, Equ. \eqref{eqn:lagposterior} and \eqref{eqn:logposteriorstrucfunc}, we then tried to recover the delay $\tau_{\mathrm{delay}}$ used for generating the light curves before.

In its left panel, Figure \ref{fig:percrec} shows a histogram of the time-intervals between SDSS S82 observational epochs of one example light curve, illustrating the possible time-delays that are covered by the data. The right hand panel shows the percentage of light curves out of this sample (10 light curves per $\tau_{\mathrm{delay}}$) for which $\tau_{\mathrm{delay}}$ can be recovered within $\pm$ 1 day. The allowed difference of 1 day was chosen to not allowing a larger absolute error for objects having larger $\tau_{\mathrm{delay}}$. Test data used for this have all the same time sampling and structure function parameters, but are light curves of different "objects". The values of $e$ and $s$ of our mock data were set to $e=0.2$, $s=1.0$. As transfer function, we used a $\delta$ function. Due to the way test data were generated, no line EW was set. (For details on test data, see \ref{sec:TestData} in the Appendix.) "Recovered" is here meant in sense of the value at maximum at posterior. A comparison of the two panels shows that time delays that match common epoch differences in S82 (left panel) can be well recovered. This makes the histogram of observational time lags a very useful tool to estimate quickly if the expected time delay should be recoverable, given the time sampling of the light curve in case. These histograms differ among the light curves in S82, as there are common time sampling windows due to the SDSS, but the exact sampling and the number of time lags being available differs.  

%% file: section6.tex
\section{RESULTS}
\label{sec:Results}
 
With the analysis tools from the previous section in place, we now proceed estimate time delays for sub-sets of the S82 data, which can then be compared to relations for $R_{\mathrm{BLR}}$ from \cite{Kaspi2000} and \cite{Bentz2013}, and relations for $M_{\mathrm{BH}}$ from \cite{Vestergaard2002}.

Given then small expected signal and the difficulties with the S82 time sampling, we found
it useful to not focus on the $\tau_{\mathrm{delay}}$ estimates of individual objects, but to presume that there is a $R_{\mathrm{BLR}}(L)$ relation, and determine its scaling normalization in different redshift (and hence luminosity) regimes, by jointly modelling several light curves.

Previous reverberation mapping studies show a simple relationship between the size of the BLR and the corresponding continuum luminosity $L$ of the form $R_{\mathrm{BLR}} \propto L^{\gamma}$ \citep{Kaspi2000}. This is an important result as it provides a secondary method of estimating on the central black hole masses by using $L$ as proxy for $R_{\mathrm{BLR}}$. This makes it a powerful tool for mass estimation in large ensembles, since a single AGN spectrum yields both $L$ and a line width $\Delta V$ suitable for estimating the size of the broad line region by using $L^{\gamma}$ and then estimating $M_{\mathrm{BH}}$ by applying Equation \eqref{eqn:massrev}. The AGN sample evaluated in this study allow us to re-address the issue of the $R_{\mathrm{BLR}}-L$ and $R_{\mathrm{BLR}}-M_{\mathrm{BH}}$ relations in AGNs.

We present novel empirical relationships for estimating the BLR sizes in AGN developed using multi-epoch photometry combined with single-epoch spectroscopy. The found scaling relationships between line widths and luminosity are based on empirical relationships between the BLR size and luminosities in various bands by \cite{Kaspi2000}, \cite{Bentz2013} and \cite{Vestergaard2002}. To obtain more definite results on the $R_{\mathrm{BLR}}-L$ and $M_{\mathrm{BH}}-L$ relations, we evaluate well-defined sub-samples of reverberation-mapped AGNs as shown in the Tables \ref{tab:sample1} to \ref{tab:sample4}.

The redshift requirements combined with the strength and $S/N$ of the emission lines make the following redshift ranges most suitable: $z=0.225-0.291$ (with 43 light curves), $z=0.555-0.591$ (with 118 light curves), $z=0.592-0.846$ (with 746 light curves).
 
Not all light curves of the 9156 spectroscopically confirmed SDSS S82 quasars (see \cite{Schmidt2010}, \cite{Schneider2007}) can be evaluated, mostly due to inappropriate time sampling with respect to the expected time delay. The expected time delay is estimated from  Kaspi relation \eqref{eqn:kaspi_r_blr_5100} if the rest-frame 5100 $\mathrm{\mathring{A}}$ luminosity is available, and from the virial assumption based on the FWHM of the H$\beta$ \eqref{eqn:virialassumption} line otherwise. We found that 35 out of the 43 light curves at $z \sim 0.25$, 69 at $z \sim 0.57$ and 290 at $z \sim 0.6 - 0.85$ have reasonable epoch coverage.

For comparing ensemble results to known mass-luminosity relations, we have to omit light curves. Specifically, light curves resulting in an unreliable posterior probability distribution were excluded from the samples. For ensemble estimates of BH masses, we have set an prior cutoff at $\tau_{\mathrm{delay}} / \tau_{\mathrm{delay, expected}} = 4$ and omitted light curve whose individual mass estimate posterior PDF is increasing towards this cutoff or is flat. We base our study on the 323 AGNs for which we can calculate reliable reverberation-based $R_{\mathrm{BLR}}$ estimates. For comparison, earlier studies used much fewer objects, e.g. \cite{Kaspi2000} based on 17 QSOs, \cite{Vestergaard2002} based on 32 AGNs, which were mostly at lower redshifts.

Tables \ref{tab:sample1} to \ref{tab:sample4} in the Appendix \ref{cha:Tables} present detailed information about the four sub-samples used for determining the $R_{\mathrm{BLR}}-L$ and $M_{\mathrm{BH}}-L$ relations and the results of individual objects. Throughout this paper, we will use the \texttt{headobjid} to identify individual objects.

\begin{figure*}
\centering
\subfigure[]{\includegraphics{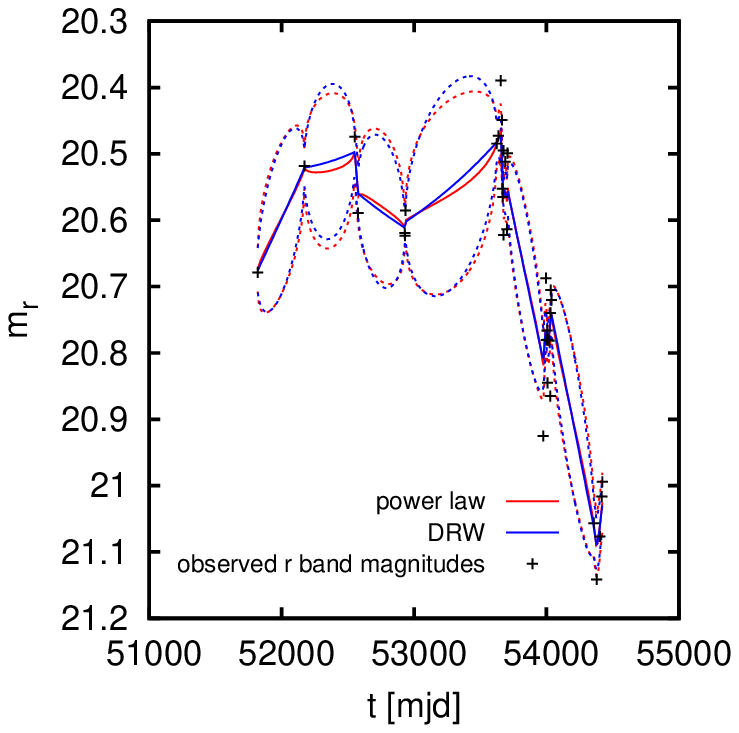}    }
\subfigure[]{\includegraphics{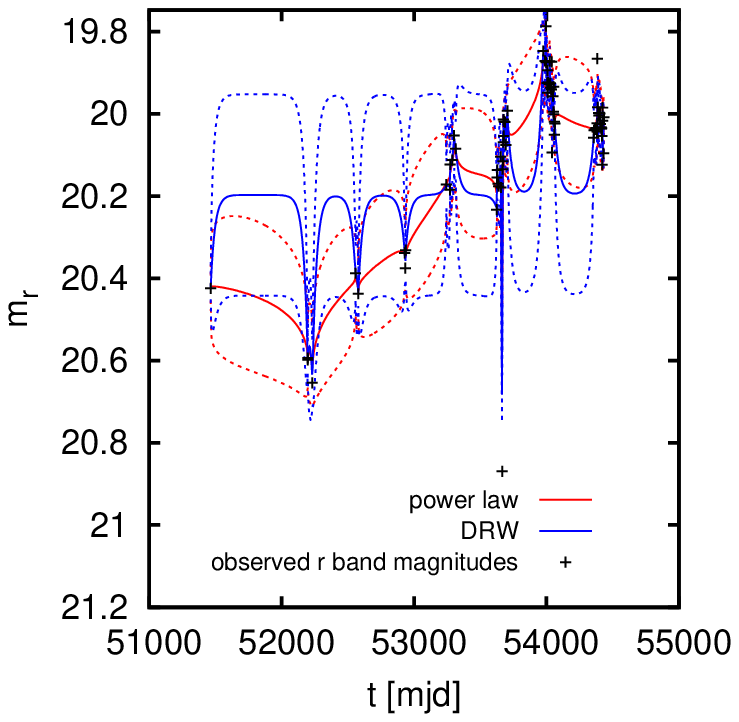}    }
\caption{Result of the interpolation of continuum light curve (\textit{r} band) for DRW and power law for a) object headobjid=587730845814686076, b) object headobjid=587731186724373007, derived from the best fit (structure function parameters at maximum at posterior) to the light curve's structure function. The solid line represents the best fit mean model light curves. The area between the dashed lines reflects the variance for the light curve prediction, arising from the stochastic models. \newline
In most cases the two functional forms lead to very similar interpolations like in a), but in a modest number of cases the DRW provides an unsatisfying fit as in b). This made lightcurve interpolation on the basis of power-law structure functions more robust for this context.}
\label{fig:fitcontinuum_powerlaw_drw_lc_587730845814686076_587731186724373007}
\end{figure*}

In the following, we present our results from using a power law structure function with posterior given by Equ. \eqref{eqn:powerlawmodelposterior} and assuming a $\delta$-function transfer function where the posterior is given by Equ. \eqref{eqn:lagposterior}. We restrict ourselves to the power law structure function, as we have found it to produce less covariance between the $\tau_{\mathrm{delay}}$ estimates and the structure function parameters. Both from real and mock data, we found that using the DRW model is not successful for application to reverberation mapping of SDSS S82 light curves. During testing (with mock light curves), it came out that fitting with the DRW lead to very unprecise estimates for $\tau_{delay}$ even with the given priors. 
The reason for this is that estimation of the fit parameters works not so well for sparsely sampled data as the fit parameters $\omega$ and $\tau$ of the 
structure function indicate the intrinsic variance of the process ($\omega^2$) and the damping timescale ($\tau$) what cannot be estimated well when having 
sparsely sampled data. However, the power law model works very good in this case, as the amplitude $A$ quantifies the root-mean-square magnitude difference on 
a one year timescale and $\gamma$ is the logarithmic gradient of this mean change in magnitude, what is easier to estimate. For a lot of SDSS S82 quasar light curves, the shape of the fitted light curve differs considerably between the power law and the DRW model, as the DRW leads to a fit that is less smooth as from the power law, and shows too much sensitivity to outliers in a lot of cases. For comparison, the power law and DRW model fits for two light curves are shown in Fig. \ref{fig:fitcontinuum_powerlaw_drw_lc_587730845814686076_587731186724373007}.

\subsection{Individual and Ensemble Estimates of BLR Sizes}

Caution must be exercised when using time lag estimates to calculate the size of broad line regions. It is the fact that for some objects, different reverberation mapping campaigns state different values for $\tau_{\mathrm{delay}}$. The implementation of corrections, e.g modeling the variation of the spectrum over time, is beyond the scope of this paper. We convert the computed observer-frame time delays $\tau_{\mathrm{delay, obs}}$ directly into BLR sizes after applying a cosmological $(1+z)^{-1}$ factor, so
\begin{equation}
R_{\mathrm{BLR}}= c \, \tau_{\mathrm{delay, obs}} (1+z)^{-1}.
\end{equation}
Individual $R_{\mathrm{BLR}}$ for the members of our sub-samples are listed in Tables \ref{tab:sample1} to \ref{tab:sample4} in the Appendix.

To define $R_{\mathrm{BLR}}$ - $L$ relations, we follow \cite{Kaspi2000} and \cite{Bentz2013}, using $\lambda L_{\lambda} (5100 \mathrm{\mathring{A}})$ as our luminosity measure. \citet{Kaspi2000} found for the BLR size-luminosity relation for the H$\alpha$ line
\begin{align} \label{eqn:kaspi_r_blr_5100}
\begin{split} 
R_{\mathrm{BLR,Kaspi}}= &\left(  32.0  ^{+2.0}_{-1.9}    \right) \left({\textstyle \frac{\lambda L_{\lambda}(5100 \; \mathrm{\mathring{A}})}{10^{44} \mathrm{ \;erg \;s^ {-1}}} }\right) ^{0.700 \pm 0.033} \\ & \textrm{light days},
\end{split}
\end{align}
which was updated by \cite{Bentz2013} as

\begin{align} \label{eqn:bentz_r_blr_5100}
\begin{split} 
R_{\mathrm{BLR,Bentz}}= & \left( 33.651  ^{+2.490}_{-2.318}    \right) \left( {\textstyle\frac{\lambda L_{\lambda}(5100 \; \mathrm{\mathring{A}})}{10^{44} \mathrm{ \;erg \;s^ {-1}}} }\right) ^{ 0.533^{+0.035}_{-0.033}} \\ & \textrm{light days}.
\end{split}
\end{align}

For the relationships, we adopt the simple form $R_{\mathrm{BLR}} \propto R_{\mathrm{BLR,Kaspi}}$ and $R_{\mathrm{BLR}} \propto R_{\mathrm{BLR,Bentz}}$, where $R_{\mathrm{BLR,Kaspi}}$ and $R_{\mathrm{BLR,Bentz}}$ are the estimates from Kaspi and Bentz, respectively. So, we do not determine a new slope but only a new proportionality constant. 

We have calculated the individual posterior probability distributions (PDFs) from evaluating Equation \eqref{eqn:lagposterior} and introducing $R_{\mathrm{BLR,Kaspi}}$ as prior using Equ. \eqref{eqn:priorgauss}. We have then projected these individual PDFs as histograms, as we marginalize over the flux scaling factors $e$ and $s$. As we assume that $\tau_{\mathrm{delay}}/\tau_{\mathrm{delay,exp}}=f(L,z,\mathrm{EW}_{\mathrm{line}})$, this marginalization was done for different cases:

For comparing to the relations \eqref{eqn:kaspi_r_blr_5100} and \eqref{eqn:bentz_r_blr_5100}, we did binning by $z$ according to our sub-sample Tables from \ref{tab:sample1} to \ref{tab:sample4}. We also did binning by luminosity $\lambda L_{\lambda}(5100 \mathrm{\mathring{A}})$. We evaluated 29 light curves in the redshift range
$z=0.225 - 0.291$, with \textit{i} band: H$\alpha$, \textit{z} band: continuum. 17 light curves out of this range were also evaluated with \textit{r} band: H$\beta$, H$\gamma$ (plus some other), \textit{g} band: continuum. We evaluated 68 light curves in the redshift range $z=0.555 - 0.591$ with \textit{g} band \ion{Mg}{2}, \textit{r}: continuum. We evaluated 111 light curves in the redshift range $z=0.592 - 0.6999$, with \textit{g} band \ion{Mg}{2} (\ion{Fe}{2} $\lambda$2600), \textit{r} continuum. The redshift range $z=0.7 - 0.846$ was evaluated with \textit{g} band \ion{Mg}{2} (\ion{Fe}{2} $\lambda$2600), \textit{r} continuum for 115 light curves.

\begin{figure*}
			 \subfigure[ ][\footnotesize{sample 1: $z$=0.225 - 0.291, with \textit{i} band: H$\alpha$, \newline \textit{z} band: continuum (29 light curves)}]  
             { \includegraphics{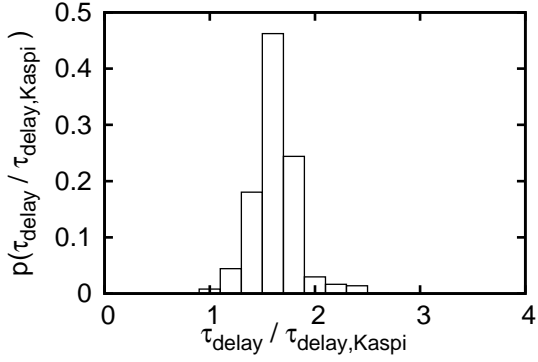}      }
			 \subfigure[ ][\footnotesize{sample 1: $z$= 0.225 - 0.291, with \textit{r} band: H$\beta$, H$\gamma$ (plus some other), \textit{g} band: continuum (17 light curves)}]       {
              \includegraphics{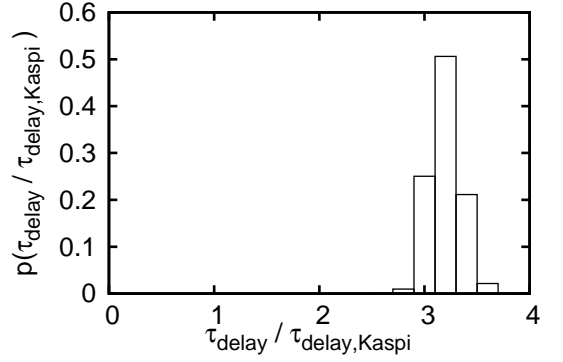}    }
			 \subfigure[ ][\footnotesize{sample 2: $z$=0.555 - 0.591, with \textit{g} band \ion{Mg}{2}, \textit{r}: continuum (68 light curves)}]        {
              \includegraphics{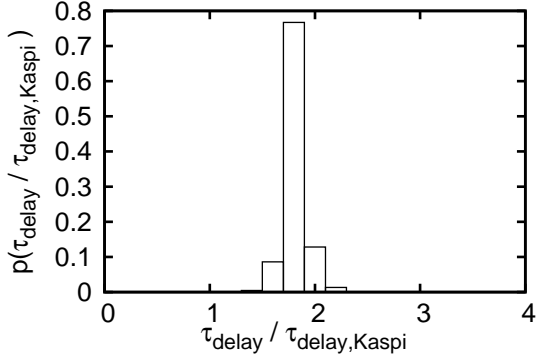}      }
			 \subfigure[ ][\footnotesize{sample 3: $z$=0.592 - 0.6999, with \textit{g} band \ion{Mg}{2} (\ion{Fe}{2} $\lambda$2600), \textit{r} continuum (111 light curves)}]        {
              \includegraphics{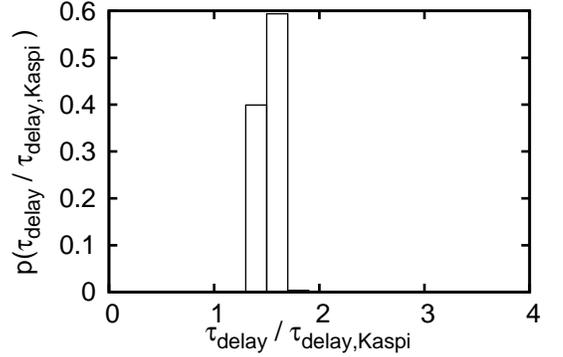}      
}
			 \subfigure[ ][\footnotesize{sample 4: $z$=0.7 - 0.846, with \textit{g} band \ion{Mg}{2} (\ion{Fe}{2} $\lambda$2600), \textit{r} continuum (115 light curves)}]        {
              \includegraphics{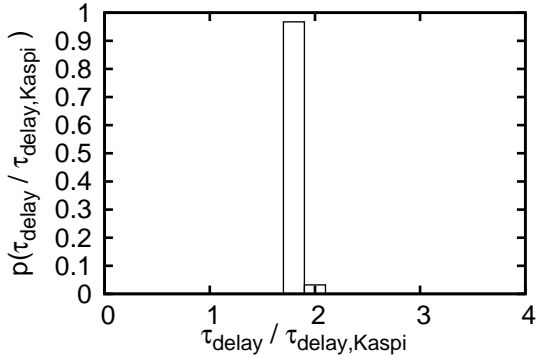}      
}
\caption[marginalized ensemble PDFs]{\footnotesize{Marginalized ensemble PDFs for comparison of the delays inferred here to those from Kaspi relation \eqref{eqn:kaspi_r_blr_5100} for different sub-samples and different emission lines. For each sub-sample, the redshift as well as the emission lines and the band used for continuum are given.}}
\label{fig:kaspi}
\end{figure*}

In Figures \ref{fig:kaspi} (a) to (e), we show the marginalized posterior probability distributions for the case of comparing the ensemble $R_{\mathrm{BLR}}$ to the Kaspi relation for $z$-binned samples. For each sample, the redshift as well as the emission lines and the band used for continuum are given.

In Figures \ref{fig:tautauexp} (a) to (d) we show the deviations of our ensemble $R_{\mathrm{BLR}}$ estimates from those of the Kaspi and Bentz relations as a function of $z$ and $L$.
Our results show ensemble estimates being about 1.7 times larger than those from the scaling relations by Kaspi and Bentz.
There is no correlation between $e$ and $\tau_{\mathrm{delay}}$. A typical shape of the probability distribution functions (PDF) is shown in Fig. \ref{fig:plotlogl_lag_elc_587731185126146081}. 

The second estimate for the first subsample of 17 light curves, done with H$\beta$ and H$\gamma$ in \textit{r} band, gives no sensible result (see Fig. \ref{fig:kaspi} (b)). Even when omitting light curves having clearly unreliable posterior PDF, PDF tends to prior cutoff at $\tau_{\mathrm{delay}} / \tau_{\mathrm{delay, expected}} = 4$. A reason for this might be that H$\beta$ is mostly weak, and there is contribution from H$\gamma$ and \ion{O}{3} $\lambda$4959, $\lambda$5007. So our assumption of having most contribution from one broadened line (here: H$\beta$) doesn't hold. For two luminosity bins in Fig. \ref{fig:tautauexp} (d), the points would lie above the prior cutoff line, so we consider them as not reliable. They are only given for reasons of completeness.

\subsection{Individual and Ensemble Estimates of BH Masses}

We convert the computed observer-frame time delays $\tau_{\mathrm{delay,obs}}$ into restframe delays applying a cosmological $(1+z)^{-1}$ factor. Then they are converted to reverberation-based $M_{\mathrm{BH}}$ by applying Equ. \eqref{eqn:massrev}. To define $M_{\mathrm{BH}}-L$ relations, we follow \cite{Vestergaard2002}, using $\lambda L_{\lambda}(5100 \mathrm{\mathring{A}})$ and $L_{\mathrm{H}\beta}$ as luminosity measure.

Vestergaard et al. \citep{Vestergaard2002,Vestergaard2006} found four empirical mass scaling relationships between line widths and luminosity for estimating $M_{\mathrm{BH}}$ in nearby AGNs and distant luminous quasars up to $z \sim 6$. In detail, they found the following relationships for the optical regime:

\begin{align}\label{eqn:vestergaard1}
\begin{split} 
\log M_{\mathrm{BH,Vestergaard1}} &= \log \left( \left( {\textstyle\frac{\mathrm{FWHM}(\mathrm{H}\beta)}{1000 \; \mathrm{km \; s^{-1}}}    }  \right)^2      \left({\textstyle \frac{\lambda L_{\lambda}(5100 \; \mathrm{\mathring{A}})}{10^{44} \mathrm{ \;erg \;s^ {-1}}} }\right)^{0.50} \right) \\& \;\; \;  + (6.91 \pm 0.02) 
\end{split}
\end{align}

\begin{align} \label{eqn:vestergaard2}
\begin{split} 
\log M_{\mathrm{BH,Vestergaard2}} &= \log \left( \left( {\textstyle\frac{\mathrm{FWHM}(\mathrm{H}\beta)}{1000 \; \mathrm{km \; s^{-1}}}    }\right)^2   
\left({\textstyle \frac{  L_{\mathrm{H}\beta}}{10^{42} \mathrm{ \;erg \;s^ {-1}}}  } \right)^{0.63} \right) \\& \;\; \; + (6.67 \pm 0.03).
\end{split}
\end{align}

Additional relationships exist for the UV, which cannot be applied here as no line widths are available for them from SDSS S82 data. They are given in \cite{Vestergaard2006}.

For comparing to the masses from the Vestergaard relations, we calculate the reverberation-mapping based $M_{\mathrm{BH}}$ by using Equ. \eqref{eqn:massrev}. We set $f=1$, and so we get
\begin{equation}
M_{\mathrm{BH,rev}}= 0.195  \, \left(\frac{\mathrm{FWHM(H \beta)} }{\mathrm{km \; s^{-1}}}\right)^2 \frac{\tau_{\mathrm{delay}}}{\mathrm{days}} \;  \mathrm{M}_{\sun}.
\end{equation}

For determining the proportionality constant to the relationships, we adopt the form $M_{\mathrm{BH}} \propto M_{\mathrm{BH,Vestergaard1}}$ and $M_{\mathrm{BH}} \propto M_{\mathrm{BH,Vestergaard2}}$. 

We have calculated the posterior probability distributions from evaluating Equation \eqref{eqn:lagposterior} projected as histograms, as we marginalize over the flux scaling factors $e$ and $s$. As we assume that $\tau_{\mathrm{delay}}/\tau_{\mathrm{delay,exp}}=f(L,z,\mathrm{EW}_{\mathrm{line}})$, this was done for different binning cases: 

For comparing to the relations \eqref{eqn:vestergaard1} and \eqref{eqn:vestergaard2}, we binned by $z$ according to our sub-sample Tables from \ref{tab:sample1} to \ref{tab:sample4}. We evaluated 29 light curves in the redshift range
$z=0.225 - 0.291$, with \textit{i} band: H$\alpha$, \textit{z} band: continuum. 17 light curves out of this range were also evaluated with \textit{r} band: H$\beta$, H$\gamma$ (plus some other), \textit{g} band: continuum. We evaluated 68 light curves in the redshift range $z=0.555 - 0.591$ with \textit{g} band \ion{Mg}{2}, \textit{r}: continuum. We evaluated 111 light curves in the redshift range $z=0.592 - 0.6999$, with \textit{g} band \ion{Mg}{2} (\ion{Fe}{2} $\lambda$2600), \textit{r} continuum. The redshift range $z=0.7 - 0.846$ was evaluated with \textit{g} band \ion{Mg}{2} (\ion{Fe}{2} $\lambda$2600), \textit{r} continuum for 115 light curves.

In Figures \ref{fig:tautauexp} (e) to (f) we show the deviations of our reverberation-based ensemble $M_{\mathrm{BH}}$ estimates from those of the Vestergaard relations as a function of $z$. Our result shows again ensemble estimates larger than those by the scaling relationships form Vestergaard by a factor of about 1.7, but with slightly larger deviations.

\subsection{Accuracy of the $R_{\mathrm{BLR}}-L$ and $M_{\mathrm{BH}}-L$ estimates}

As we have illustrated in Figure \ref{fig:percrec}, time delays can only be robustly recovered if the fall into certain windows that are set by the S82 sampling. We now check
the time-delays inferred for the S82 from our analysis {\it post facto} against this criterion.

\begin{figure*}
			 \subfigure[ ]        {
            \includegraphics{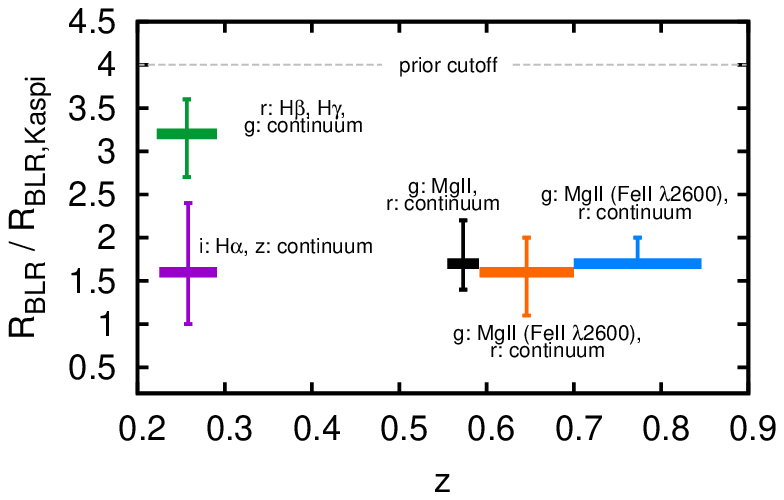}      
             }      \qquad
			 \subfigure[ ]
            {
              \includegraphics{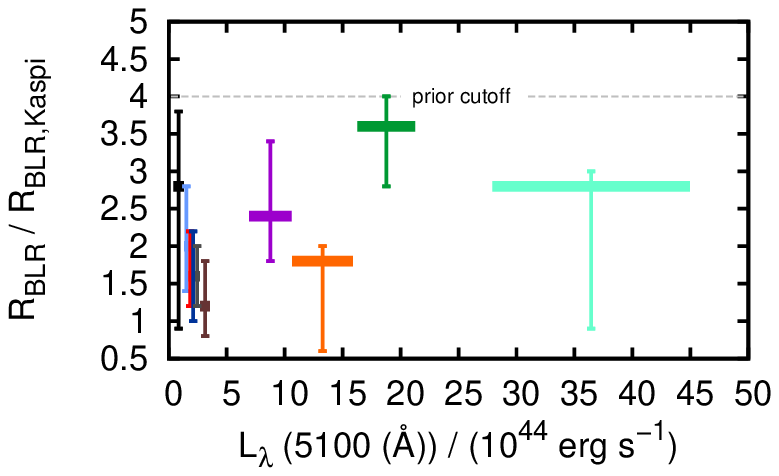}   
            }  \\ 
             \subfigure[ ]
             {
              \includegraphics{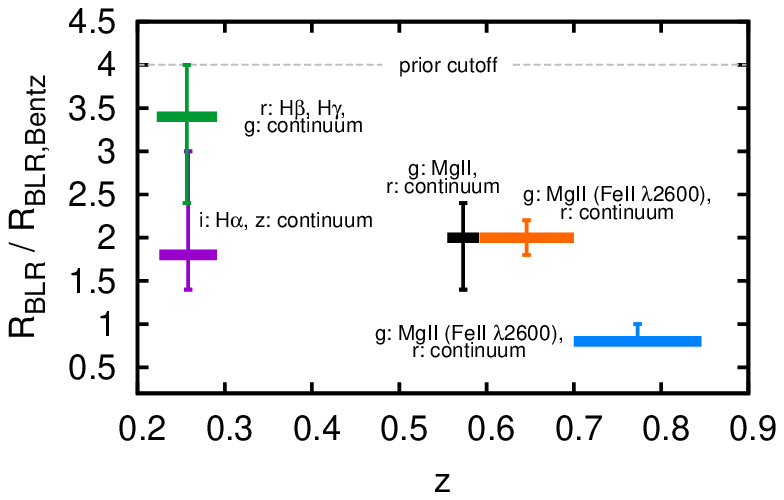}    
             }   \qquad   
             \subfigure[ ]
             {
              \includegraphics{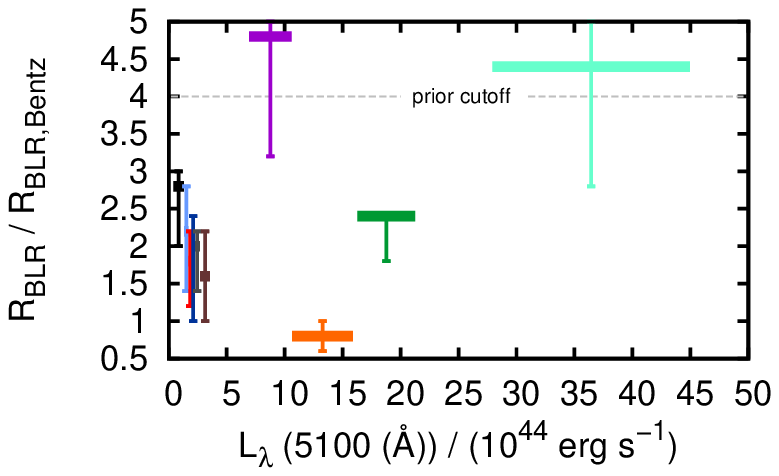}       
             }   \\ 
			 \subfigure[ ]
             {
              \includegraphics{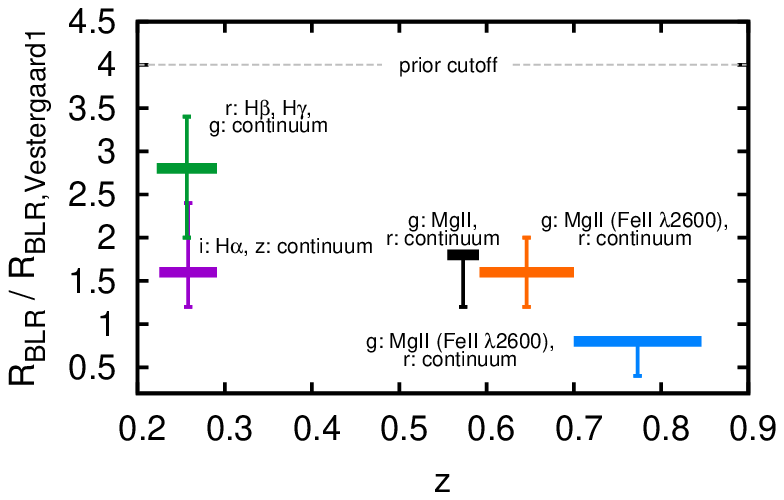}    
             }    \qquad       
			 \subfigure[ ]
             {
              \includegraphics{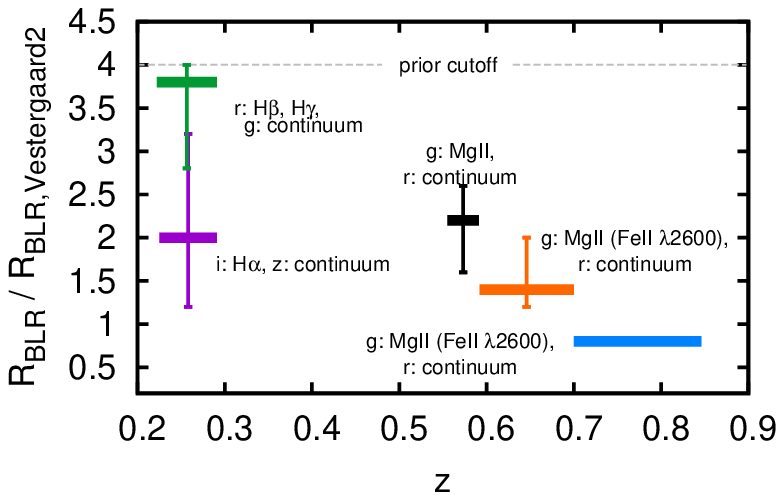}     
             }   
             
\caption[comparing our results to \cite{Kaspi2000}, \cite{Bentz2013}, \cite{Vestergaard2002}]{\footnotesize{Comparing our results to \cite{Kaspi2000}, \cite{Bentz2013}, \cite{Vestergaard2002} \newline 
(a) ensemble $R_{\mathrm{BLR}} /R_{\mathrm{BLR,Kaspi}}$ binned by redshift\newline 
(b) ensemble $R_{\mathrm{BLR}} /R_{\mathrm{BLR,Kaspi}}$ binned by luminosity \newline 
(c) ensemble $R_{\mathrm{BLR}} /R_{\mathrm{BLR,Bentz}}$ binned by redshift\newline 
(d) ensemble $R_{\mathrm{BLR}} /R_{\mathrm{BLR,Bentz}}$ binned by luminosity \newline 
(e) ensemble $M_{\mathrm{BH}} /M_{\mathrm{BH,Vestergaard1}}$ binned by redshift\newline 
(f) ensemble $M_{\mathrm{BH}} /M_{\mathrm{BH,Vestergaard2}}$ binned by redshift.\newline
For the sake of clarity, bins are colored. Each color is indicating the same bin for all diagrams showing binning by $z$. Each color is indicating the same bin for all diagrams showing binning by luminostiy. For $z$ bins, also the used bands and their broad emission lines along with contaminating lines are given.
}}
\label{fig:tautauexp}
\end{figure*}

\begin{figure*}[H]
			 \subfigure[ ]        
			 {
            \includegraphics{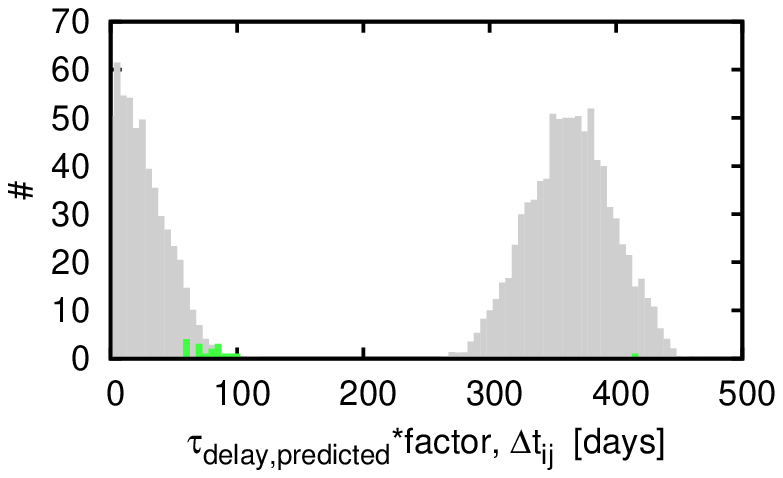}      \qquad
             }      
			 \subfigure[ ]
            {
              \includegraphics{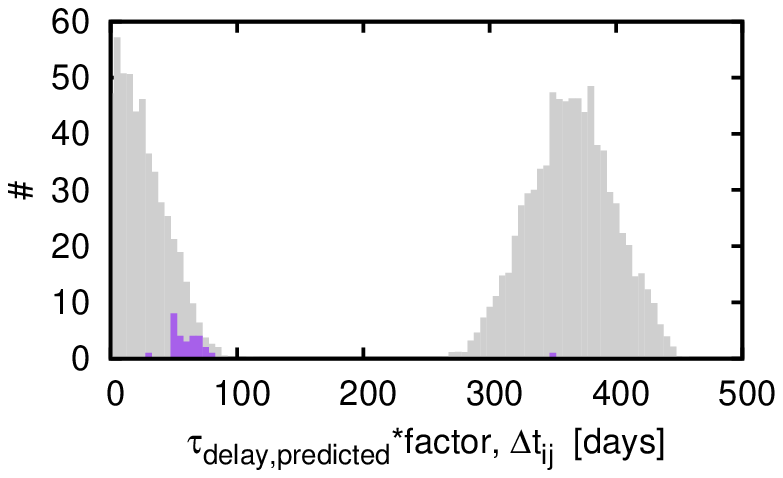}    \\  
            }
             \subfigure[ ]
             {
              \includegraphics{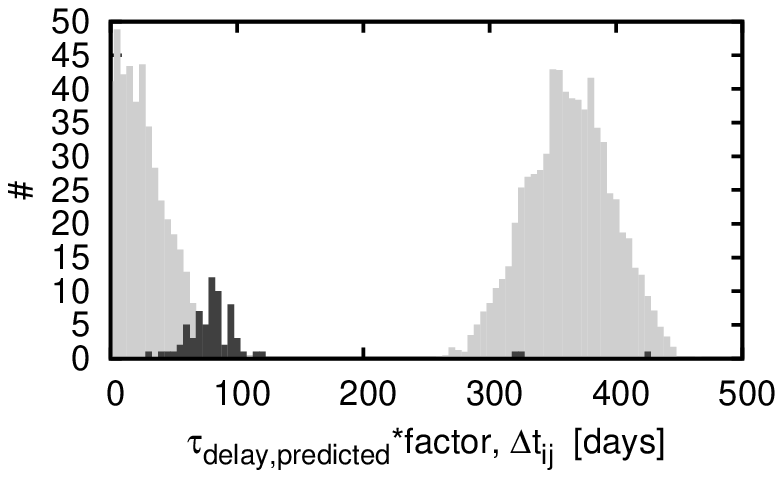}   \qquad    
             }   
             \subfigure[ ]
             {
              \includegraphics{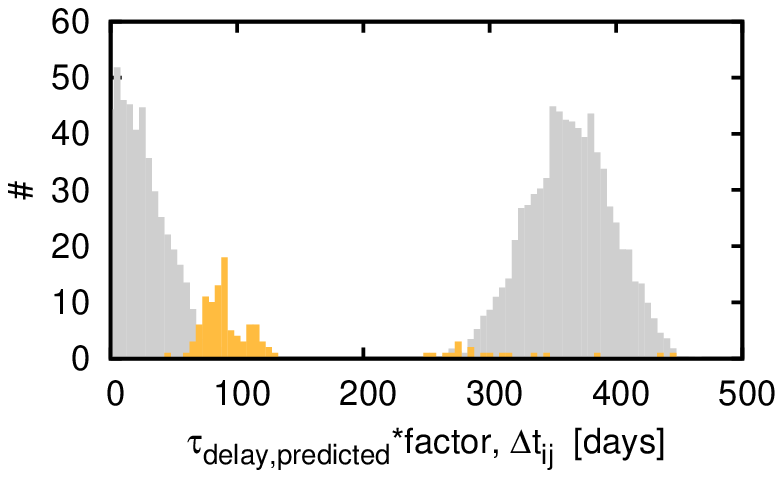}     \\    
             } 
			 \subfigure[ ]
             {
              \includegraphics{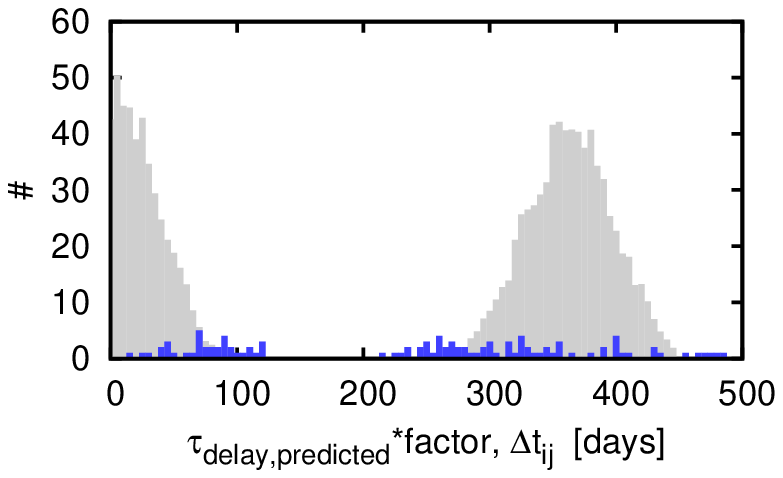}    \qquad 
             }        
\caption{\footnotesize{The grey histograms indicate the ensemble-mean observational time lag distribution for the light curves in each redshift bin used to calculate the ensemble relation; this histogram indicates (see Fig. \ref{fig:percrec}) the regime for $\tau_{\mathrm{delay}}$ where its value can be robustly recovered. The various colored histograms indicate the distribution of the predicted $\tau_{\mathrm{delay}}$ from Kaspi relation scaled by the proportionality factor as we inferred from the data (see Fig. \ref{fig:tautauexp}(a)) in the different redshift bins. The comparison of the grey and colored histograms shows that, except for perhaps (b) and (e) the S82 sampling is expected to affect seriously the robustness of the formally inferred time delays.
 \newline 
(a) \textit{r}: H$\beta$, H$\gamma$, \textit{g}: continuum, (b) \textit{i}: H$\alpha$, \textit{z}: continuum  \newline 
(c) \textit{g}: \ion{Mg}{2}, \textit{r}: continuum, (d) \textit{g}: \ion{Mg}{2} (\ion{Fe}{2} $\lambda$2600), \textit{r}: continuum \newline 
(e) \textit{g}: \ion{Mg}{2} (\ion{Fe}{2} $\lambda$2600), \textit{r}: continuum }}
\label{fig:tau_delay_predicted_kaspi_observationlaghistogram}
\end{figure*}

For this purpose we made a mean histogram of the inferred ensemble time lags for all the 323 light curves used to calculate the ensemble relation (Fig. \ref{fig:tautauexp}). For each of these objects, we used their UV luminosity and linewidth in conjunction with the Kaspi relation -- scaled by the factor from Fig. \ref{fig:tautauexp}(a) for objects in any given bin --, to predict their most likely $\tau_{\mathrm{delay}}$. This procedure results in a histogram of the inferred time delay $\tau_{\mathrm{delay,predicted}} \times factor$ for the objects in each bin, which is shown as colored histograms in the panels of Fig. \ref{fig:tau_delay_predicted_kaspi_observationlaghistogram}. Specifically, the scaling factors we used were the most likely ones for each bin, e.g., 3.2 for Fig. \ref{fig:tau_delay_predicted_kaspi_observationlaghistogram}(a), 1.6 for Fig. \ref{fig:tau_delay_predicted_kaspi_observationlaghistogram}(b), 1.7 for Fig. \ref{fig:tau_delay_predicted_kaspi_observationlaghistogram}(c), 1.6 for Fig. \ref{fig:tau_delay_predicted_kaspi_observationlaghistogram}(d), 1.7 for Fig. \ref{fig:tau_delay_predicted_kaspi_observationlaghistogram}(e).

Presuming Kaspi relation with factor 1 were true, the $\tau_{\mathrm{delay}}$ should fall into a region where we can recover them, as this was one of the selection criteria. 
If the inferred scaling factors from our analysis were true, the implied delays fall into the tails of the distribution. In this cases, correct inference may occur, but its recovery robustness is not particularly likely. From further analysis, most of the individual $\tau_{\mathrm{delay}}$ fall in regions were less than 50 percent of the time delays are assumed to be calculated correctly.

Additionally, reverberation-based masses are themselves typically uncertain by a factor of $f =2-3$. The absolute accuracy of the single-epoch mass estimates by \cite{Vestergaard2002} is stated to be between a factor of 3.6 and 4.6. Also, one has to keep in mind that the scaling relations and their uncertainties are of statistical nature. Any given single estimate from a $R_{\mathrm{BLR}}-L$ or $M_{\mathrm{BH}}-L$ scaling relation can be off by some factor. Therefore it should not be trusted in cases where high accuracy is needed. Such relations are, however, a useful tool for application to large statistical samples.

%% file: section7.tex
\section{CONCLUSION AND DISCUSSION}
\label{sec:ConclusionAndDiscussion}

For the purpose of evaluating sparsely sampled photometric data, we implemented an advanced stochastical reverberation mapping algorithm in order to find correlated variations in a purely continuum and a continuum plus emission line band. This method is based on an approach for spectroscopic reverberation mapping by \cite{Rybicki1994} and \cite{Zu2011} and extended for being capable of handling sparsely sampled multi-epoch photometric data in combination with constraining single-epoch spectroscopy. This enables us to use data of available long-term photometric surveys, where we explored whether photometric reverberation mapping is feasible. Having a method to evaluate such data enables us to apply reverberation mapping for the first time to large samples of a few hundred AGN at far higher redshifts than before.

We have set out to obtain individual and ensemble estimates of the BLR size in quasars, and on that basis explore their $M_{\mathrm{BH}}$. We did this by means of photometric reverberation mapping, drawing on existing data in SDSS Stripe 82. Our results show the power of stochastical broad-band reverberation mapping techniques of quasar ensembles, in contrast to "classical" spectroscopic monitoring of individual objects. This enables us to make use of available long-term surveys. We present novel relationships between spectrophotometric parameters allowing one to calculate ensemble estimates of AGN central black hole masses, computed by an improved reverberation mapping method and carried out on S82 for the first time.

We model the continuum in one band as stochastic Gaussian process and assume a flux model that describes the other band with emission line contribution as a scaled version of the pure continuum band plus a smoothed and displaced version of the continuum that was generated using a $\delta$ function transfer function. Evaluation is carried out by stochastic methods. This approach can not only interpolate between data points, but also make and include self-consistently estimates, where statistical confidence limits on all estimated parameters are determined. It also is able to derive simultaneously the lags of multiple emission lines.

By generating and evaluating extensive sets of of problem-specific mock data, we made sure that our model will sufficiently fit the continuum light curve and in a second step the continuum plus emission light curve in order to solve for the time delay. We found a method to estimate if time lags in the range of the expected one can be determined in principle from a given light curve with respect to its time sampling. This is needed for making sure that we can trust in our estimates. Also, this reduces computation time a lot.

In the application to SDSS S82 data, we combine on average more than 60 epochs of photometric data with one epoch of spectroscopy. In all bandpasses the accretion disk emission (the continuum) contributes much of the flux, but in some bands, the BLR line flux contributes up to 15 $\%$. Constraining spectroscopic broad-band information is provided for the H$\alpha$, H$\beta$ and \ion{Mg}{2} lines.

We have identified the SDSS S82 temporal sampling windows as a serious issue. Despite the stochastic approach is very good in dealing with uneven and sparsely time sampling, the sampling windows make it necessary to pre-select sufficient light curves before evaluating them. This is needed as data usually exist only for 2--3 month within each year. Also, we identified suitable redshift ranges to get at minimum one band that is emission-line free and one having contribution from H$\alpha$, H$\beta$ or \ion{Mg}{2}.
For SDSS-like mock data, we were able to show that the method should work for application to individual AGN, so we can trust in our estimates for $\tau_{\mathrm{delay}}$ from SDSS S82 light curves.

We estimated the time delay $\tau_{\mathrm{delay}}$ for a well-defined sample of 323 objects spanning redshifts from $z=0.225-0.846$. In detail, we evaluated 29 light curves in the redshift range $z=0.225 - 0.291$, with \textit{i} band: H$\alpha$, \textit{z} band: continuum. We evaluated 68 light curves in the redshift range $z=0.555 - 0.591$ with \textit{g} band \ion{Mg}{2}, \textit{r}: continuum. We evaluated 111 light curves in the redshift range $z=0.592 - 0.6999$, with \textit{g} band \ion{Mg}{2} (\ion{Fe}{2} $\lambda$2600), \textit{r} continuum. In the redshift range $z=0.7 - 0.846$, we evaluated 115 light curves with \textit{g} band \ion{Mg}{2} (\ion{Fe}{2} $\lambda$2600), \textit{r} continuum. The Tables \ref{tab:sample1} to \ref{tab:sample4} in the Appendix show detailed information about the four sub-samples and the results of individual objects.

17 light curves out of the redshift range $z=0.225 - 0.291$ were also evaluated with \textit{r} band: H$\beta$, H$\gamma$ (plus some other), \textit{g} band: continuum. From this, we get no sensible results. Even when omitting light curves having clearly unreliable posterior PDF, the PDF tends to the prior cutoff at $\tau_{\mathrm{delay}} / \tau_{\mathrm{delay, expected}} = 4$. Also, it is troubling that the H$\beta$ BLR comes out larger than the H$\alpha$ BLR, in light of other results \citep{Bentz2010}. A reason for this might be the mostly weak H$\beta$ together with contribution from H$\gamma$ and \ion{O}{3} $\lambda$4959, $\lambda$5007, so our assumption of having most contribution from H$\beta$ doesn't hold.

The posterior distribution functions of the fit parameters for ensembles of these objects where evaluated by multiplying them for sensible redshift or luminosity bins. As a result, we get an ensemble scaling relationship for the scaling of the BLR size as a function of luminosity and redshift. We are also able to give constrains on the scaling relationship between the central black hole masses and the luminosity of the AGN. Comparing our results with comparable published data by \cite{Kaspi2000}, \cite{Vestergaard2002} and \cite{Bentz2013}, we find that formally the proportionality constant is significantly bigger than those published before, but S82 data makes the result, and its implication, tentative.

It is important to note that the $L-R_{\mathrm{BLR}}$ and $L-M_{\mathrm{BH}}$ relationships and their uncertainties quoted stem from the ensemble average of many objects. The uncertainties associated with individual $M_{\mathrm{BH}}$ estimates may be considerably higher.

We have shown that the robustness of our method is limited by time sampling. Presuming Kaspi relation with factor 1 were true, the $\tau_{\mathrm{delay}}$ should fall into a region where we can recover them, as this was one of the selection criteria. Assuming the Kaspi relation has to be scaled by a factor - as our results indicate -, the inferred $\tau_{\mathrm{delay}}$ are not longer falling into regions of the time sampling that are well covered. In such cases, correct inference may occur due to the interpolation, but cannot be trusted. From further analysis, most of the individual $\tau_{\mathrm{delay}}$ fall in regions were less than 50 percent of the time delays are assumed to be calculated correctly.

The formalism developed here should be useful for application to future data sets.

As we use constraining single-epoch spectra, an extension to apply the method to some more spectra for given objects may be interesting. For some AGN, SDSS provides multi-epoch spectra, where we have seen that their emission lines vary significantly in some cases.

Another useful extension would be to add the information which light-curve properties lead to good $\tau_{\mathrm{delay}}$ estimate and give different weights on them for the ensemble PDF, instead of excluding some and weight the remaining all equally.

It would be interesting to introduce a wider range of transfer functions. We have derived the formalism for dealing with not only a $\delta$ function but also a Gaussian. But the Gaussian transfer function couldn't be applied here as it cannot be applied easily to the power law structure function model for reasons of non-analytical integrals. For application to a DRW, we get analytical integrals but the DRW doesn't work well for photometric reverberation of SDSS S82 light curves. For this reason, implementation of additional transfer functions like the top-hat or sawtooth function, as used by some authors, would be useful.\newline
We know that not all parts of an emission line vary. To deal with this issue, we use the information on the broad emission line part provided by the \textit{Catalog of Quasar Properties}, in detail, the line luminosity, FWHM and restframe equivalent with of the broad emission lines. It might be feasible to include the fact that not all parts of an emission line vary in a better way as up to now in the prediction of equivalent widths from given $(e,s)$ scaling factors for calculating $\mathscr{L}_{\mathrm{spec}}$. This should be a minor effect, though, as the line profile looks very similar between the mean spectrum and the RMS spectrum, where the RMS is over spectra taken at different times \citep{Kaspi2000}.

Whereas the \textit{Catalog of Quasar Properties} provides information on the broad component of H$\alpha$, H$\beta$, \ion{Mg}{3}, and we use them, for \ion{C}{4}, only information on the whole line is available. As no information on the FWHM of the broad component of \ion{C}{4} is available from the catalog \citep{Shen2011}, no comparison to the third Vestergaard relation (see Equ. 7 in \cite{Vestergaard2006}) could be done. This might be carried out with data from upcoming surveys.

%% file: Appendix.tex
\section{THE STRUCTURE FUNCTION AND THE GAUSSIAN PROCESS}
\label{sec:TheStructureFunctionAndTheGaussianProcess}

Imagine a set of $N$ measurements $m_i$, being calibrated magnitude or flux measurements all taken in a single bandpass of a single source. Each measurement $m_i$ is taken at a time $t_i$, and has a (presumed known) uncertainty variance $\sigma_i$. The structure function $\mathrm{V}(|\Delta t|)$ is defined as follows: The expectation value $\mathrm{E}[\cdot]$ for the difference between observation $m_i$ and $m_j$ (with $i \neq j$) is
\begin{equation}
 \mathrm{E}[(m_i - m_j)^2]=\sigma_i^2+\sigma_j^2 + \mathrm{V}(|t_i-t_j|),
\end{equation}
where the observations are presumed to be (from a measurement noise perspective) independent, and the structure function $\mathrm{V}(\cdot)$ effectively describes the variance. In the literature, the structure function has occasionally been defined in terms of the root-mean square (the square root of the above definition) and sometimes in terms of the mean absolute differences, with is slightly different again.

A Gaussian process is characterized by a function describing the mean measurement $\bar{\mathbf{m}}(t)$ (magnitude or flux) as a function of time $t$ and a function $C(t,t')$ describing the covariance between observations $\mathbf{m}$ at different epochs $t$ and $t'$. We will assume that the mean is constant and that the process is stationary such that $C(t, t') \equiv C (t-t')$. The probability of a set of $N$ observations $\left\{ m_i\right \} ^N _{i=1}$ is given by that of the $N$-dimensional Gaussian with mean $\bar{\mathbf{m}}=(m, m, ..., m)^{\mathrm{T}}$ and $N \times N$ dimensional covariance matrix $C$ with elements $C_{ij}=C(t_i - t_j)$.\newline
It is possible to define a Gaussian process that generates data in accordance with any (reasonable) structure function. As the structure function is the expectation of the squared measurement differences between observations $t_i$ and $t_j$ separated by a time $\Delta t$, we can write it as \citep{Rybicki1992}
\begin{align}
\mathrm{V}(|\Delta t|)&=\mathrm{E}[(m(t)-m(t+\Delta t))^2] \notag \\ &= 2\mathrm{E}[(m(t)-\mathrm{E}[m])^2]-2\mathrm{E}[(m(t)-\mathrm{E}[m])(m(t+\Delta t)-\mathrm{E}[m]]
\end{align}
and
\begin{align}
\mathrm{V}_{ij} & \equiv  \frac{1}{2} \mathrm{V}(|t_i-t_j|) \\
\mathrm{V}_{\infty} & = \frac{1}{2}\mathrm{V}(\Delta t \rightarrow \infty).
\end{align}
The covariance function of the Gaussian process corresponding to the structure function $\mathrm{V}$ with parameters $\mathbf{p}$ is then given by 
$C_{ij}= \mathrm{V}_{\infty}-\mathrm{V}_{ij}$, or expanded
\begin{equation}
C = C(t,\mathbf{p},\sigma) =
\begin{pmatrix} 
(\mathrm{V}_{\infty}+\sigma_1^2) & (\mathrm{V}_{\infty}-V_{12}) & (\mathrm{V}_{\infty}-V_{13}) & \dotsc & (\mathrm{V}_{\infty}-\mathrm{V}_{1N})\\ 
(\mathrm{V}_{\infty}-\mathrm{V}_{21}) & (\mathrm{V}_{\infty}+\sigma_2^2) & (\mathrm{V}_{\infty}-\mathrm{V}_{23}) & \dotsc & (\mathrm{V}_{\infty}-\mathrm{V}_{2N})\\ 
(\mathrm{V}_{\infty}-\mathrm{V}_{31}) & (\mathrm{V}_{\infty}-\mathrm{V}_{32}) & (\mathrm{V}_{\infty}+\sigma_3^2) & \dotsc & (\mathrm{V}_{\infty}-\mathrm{V}_{3N})\\ 
\dotsc & \dotsc & \dotsc & \dotsc & \dotsc \\ 
(\mathrm{V}_{\infty}-\mathrm{V}_{N1}) & (\mathrm{V}_{\infty}-\mathrm{V}_{N2}) & (\mathrm{V}_{\infty}-\mathrm{V}_{N3}) & \dotsc & (\mathrm{V}_{\infty}+\sigma_N^2) 
\end{pmatrix}.
\label{eqn:covmatrix}
\end{equation}
To get the effective (observed) variability, the photometric uncertainty variances $\sigma_i$ are added to the diagonal elements of $C$.\newline
Variability models can therefore be expressed either in terms of the variance function or equivalently in terms of the structure function. For example, imagine that the two quantities $m_i$ and $m_j$ are not observations of a quasar but instead random numbers drawn from an $N$-dimensional Gaussian,
\begin{equation}
p(\mathbf{m})= \mathscr{N}(\mathbf{m}|\bar{\mathbf{m}}, C),
\end{equation}
\begin{equation}
\mathbf{m}=(m_1, m_2, m_3, ..., m_N)^{\mathrm{T}}, \; \; \bar{\mathbf{m}}= (\bar{m}, \bar{m}, \bar{m}, ..., \bar{m})^{\mathrm{T}}
\end{equation}
where we have assembled the observations into a column vector $\mathbf{m}$. $\mathscr{N}(\cdot|\bar{\mathbf{m}}, C)$ is the general normal or Gaussian PDF given mean vector $\bar{\mathbf{m}}$ and a variance tensor $C$, $\bar{m}$ is an arbitrary parameter, $2V_{ij}$ is the structure function evaluated at time lag $|t_i - t_j|$ as defined above. If we make many draws from this Gaussian, the expectation values of $(m_i-m_j)$ and $(m_i-m_j)^2$ for any pair of measurements $m_i$ and $m_j$ (with $i \neq j$) are just
\begin{equation}
\mathrm{E}[(m_i-m_j)]=0
\end{equation}
and
 \begin{equation}
 \mathrm{E}\left [(m_i-m_j)^2 \right]  = \sigma_i^2+\sigma_j^2+\mathrm{V}(|t_i-t_j|),
 \end{equation}
which, by design, is equivalent to the description of the structure function.

Two additional points arise from this description. Although $m$ does not enter in the prediction of the mean or variance of the magnitude differences, it does, of course, affect the magnitudes themselves. So it is, in principle, an observational property of the model. Although $\mathrm{V}_{\infty}$ is not, in practice, measurable, it can be approximated by evaluating the structure function at large time lag.

\subsection{The Basic Stochastic Approach}
\label{sec:TheBasicStochasticApproach}

The idea behind this approach was developed by \cite{Press1992}, \cite{Rybicki1992} and \cite{Rybicki1994} in order to estimate the best-fit structure function parameters for a given light curve. Later on, this was extended by \cite{Zu2011} and \cite{Chelouche2012} to a method that enables both structure function parameter estimation and estimation of the time delay between multiple uneven sampled light curves. Here, we refer to the method from \cite{Zu2011} and re-summarize some of its formalism. We present how we have improved it to do broad-band reverberation mapping, supported by one epoch of spectroscopy to separate continuum and emission line contribution, as well as their application.

The intrinsic variability has a covariance matrix $S= \langle \mathbf{s}\mathbf{s} \rangle$, whereas the noise has a covariance matrix $N= \langle \mathbf{n}\mathbf{n}\rangle$.
By definition, we know that
\begin{equation}
P(\mathbf{s}) \propto |S|^{-1/2} \exp \left( - \frac{\mathbf{s}^{\mathrm{T}} S^{-1}\mathbf{s}}{2}   \right)
\end{equation}
and that 
\begin{equation}
P(\mathbf{n}) \propto |N|^{-1/2} \exp \left( - \frac{\mathbf{n}^{\mathrm{T}} N^{-1}\mathbf{n}}{2}   \right).
\end{equation}
Thus, the probability of the data given the linear coefficients $\mathbf{q}$, the intrinsic light curve $\mathbf{s}$, and any other parameters of the light curve model $\mathbf{p}$ (the structure function parameters) is
\begin{equation}
P(\mathbf{m}|\mathbf{q},\mathbf{s},\mathbf{p}) \propto |SN|^{-1/2} \int  \mathrm{d}^n \mathbf{n} S(\mathbf{m}-(\mathbf{s}+\mathbf{n}+L\mathbf{q}))    \exp \left( - \frac{\mathbf{s}^{\mathrm{T}} S^{-1}\mathbf{s} + \mathbf{n}^{\mathrm{T}} N^{-1}\mathbf{n}}{2}   \right).
\end{equation}
After evaluating the Dirac $\delta$ function, we complete the squares in the exponential with respect to both the unknown intrinsic source variability $\mathbf{s}$ and the linear coefficients $\mathbf{q}$.

This determines our best estimate for the mean light curve, 
\begin{equation}
\hat{\mathbf{p}}=SC^{-1}(\mathbf{m}-L\hat{\mathbf{q}})
\label{eqn:bestestimatemeanlc}
\end{equation}
with linear coefficients
\begin{equation}
\hat{\mathbf{q}}=(L^{\mathrm{T}} C^{-1}L)^{-1} L^{\mathrm{T}} C^{-1}\mathbf{m} \equiv C_q L^{\mathrm{T}} C^{-1}\mathbf{m}.
\end{equation}
$C=S+N$ is the overall covariance matrix of the data and $C_q \equiv (L^{\mathrm{T}} C^{-1} L)^ {-1}$.\newline
With these definitions, we can factor the argument of the exponential into
\begin{equation}
P(\mathbf{m}|\mathbf{q},\mathbf{s},\mathbf{p}) \propto |SN|^{-1/2} \exp \left( - \frac{\Delta \mathbf{s}^{\mathrm{T}} (S^{-1}+N^{-1}) \Delta \mathbf{s}}{2} - \frac{\Delta \mathbf{q}^{\mathrm{T}} C_q^{-1} \Delta \mathbf{q}}{2} -\frac{\mathbf{m}^{\mathrm{T}} C_{\perp}^{-1} \mathbf{m}}{2} \right),
\end{equation}
where 
\begin{equation}
C_{\perp}^{-1} \equiv C^{-1} - C^{-1}L C_q L^{\mathrm{T}} C^{-1}
\end{equation}
is the component of the covariance matrix $C$ that is orthogonal to the fitted linear functions. The variances in the linear parameters are 
\begin{align}
\langle \Delta \mathbf{q}^2 \rangle &= (L^{\mathrm{T}} C^{-1} L)^{-1} \equiv C_q, \\
\Delta \mathbf{s} &=\mathbf{s}-\hat{\mathbf{s}}, \\
\Delta \mathbf{q} &=\mathbf{q}-\hat{\mathbf{q}} .
\end{align}
\newline
We are now prepared to marginalize the probability over the light curve $\mathbf{s}$ and the linear parameters $\mathbf{q}$ under the assumption of uniform priors for these variables. When doing so, we find that
\begin{equation}
P(\mathbf{m}|\mathbf{p}) \propto \mathcal{L}(\mathbf{m}|\mathbf{p}) \equiv |S+N|^{-1/2} |L^{\mathrm{T}} C^{-1} L|^{-1/2} \exp \left( - \frac{\mathbf{m}^{\mathrm{T}} C_{\perp}^{-1}  \mathbf{m}}{2}  \right),
\label{eqn:parameterlikelihood}
\end{equation}
where for the exponential model the remaining parameters $\mathbf{p}$ are $\tau$ and $\omega$ and for the power-law model $A$ and $\gamma$. $\mathcal{L}$ represents the likelihood function we are to maximize in order to find the most likely combination of those parameters. \newline

Mathematically, the mean light curve is the weighted average of all process light curves described by the parameter vector $\mathbf{p}$ being statistically consistent with the data, and the variance is the scatter of these light curves about this mean.

\begin{figure}[!ht]
\centering         
   \subfigure[ ]{\includegraphics{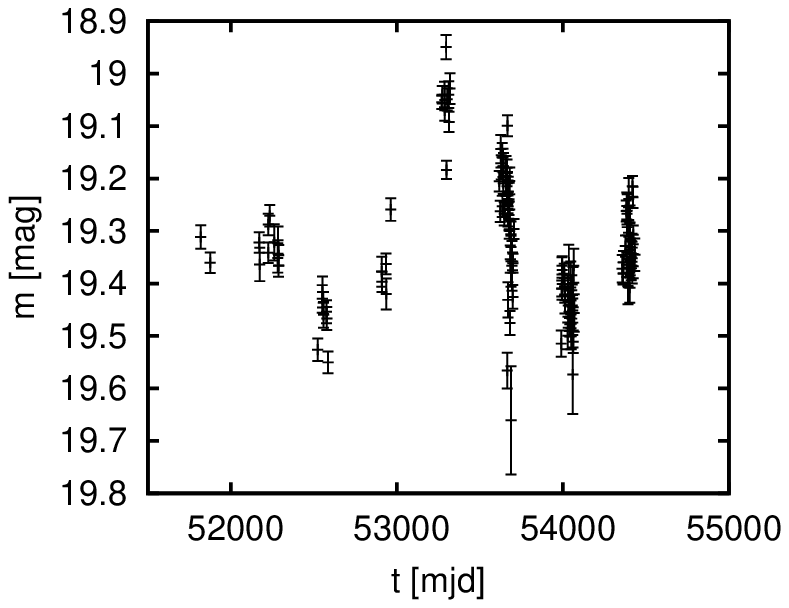}} 
   \subfigure[ ]{\includegraphics{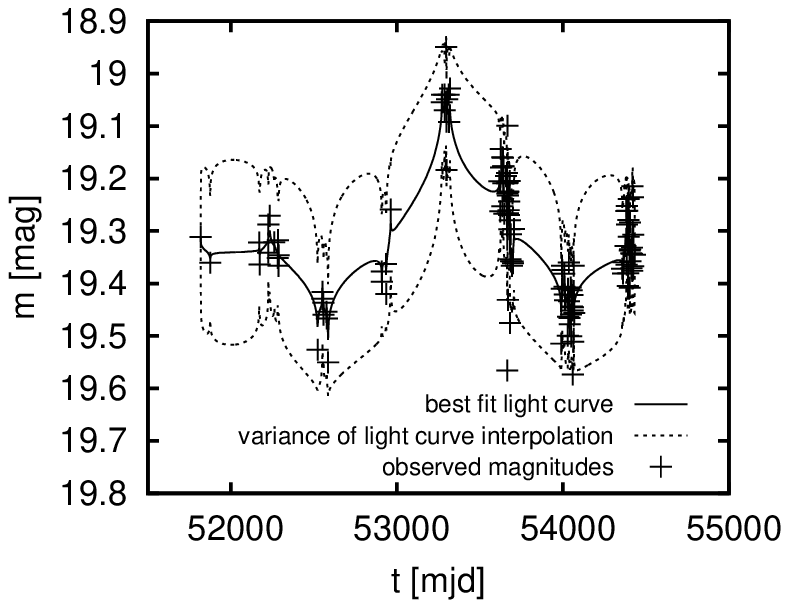}}\\
      \subfigure[ ]{\includegraphics{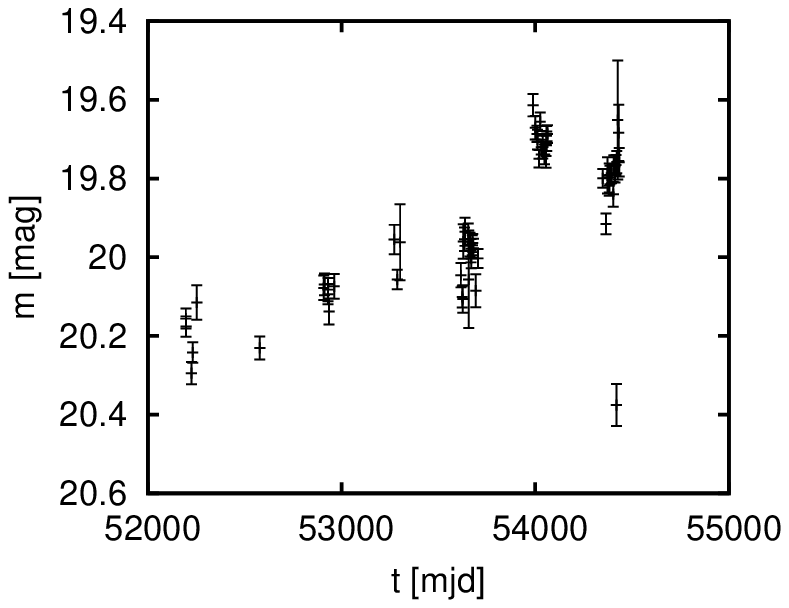}} 
   \subfigure[ ]{\includegraphics{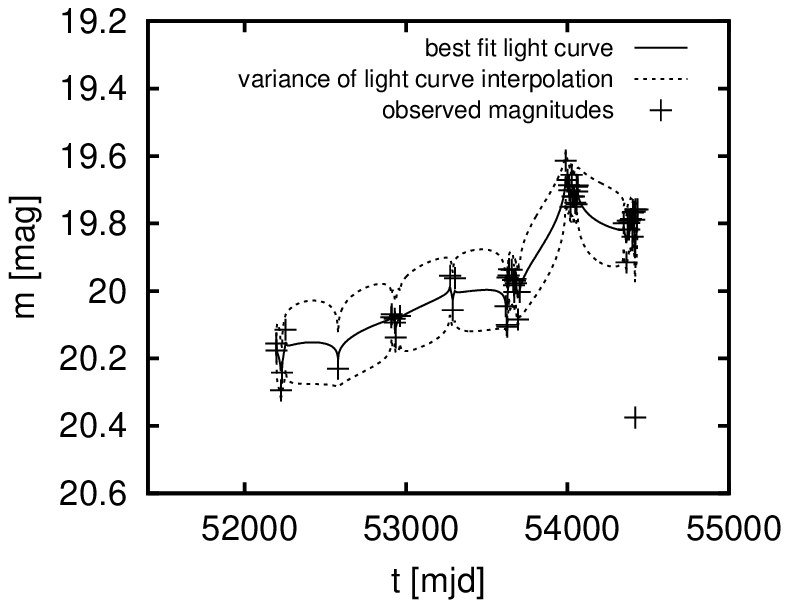}}\\
  \caption[examples of light curve models for two quasar light curves]{\footnotesize{Examples of light curve models for two quasar light curves. The light curves are from spectroscopically confirmed quasars of the SDSS Stripe 82 in a redshift region where $r$ band is only continuum. They are showing different seasonal gaps.
For fitting the light curves, outliers are excluded. The solid lines in the right figure panel represent the best fit mean model light curves from the power law model. The area between the dotted lines represents the "error snake", the 1$\sigma$ range of possible stochastic models. The "error snakes" bound the reconstructed light curve are thinner than the data points because of the additional measurement error on the data. We also give the best model parameter values along with the confidence intervals (CI). \newline
first row: SDSS S82 $r$ band quasar light curve headobjid=588015509285437517, observation period: 30.9.2000 - 28.11.2007 (2614.9 days), fitted with a $\bar{m}$=19.331 mag, $A$=0.136 (0.68 CI [0.113,0.165], (0.95 CI [0.099,0.216]), $\gamma$=0.205 (0.68 CI [0.101,0.284], (0.95 CI [0.089,0.393]) \newline
second row: SDSS S82 $r$ band quasar light curve headobjid=587731185661640908, observation period: 13.10.2001 - 28.11.2007 (2236.9 days), fitted with a $\bar{m}$=19.993 mag, $A$=0.175 (0.68 CI [0.133,0.234], (0.95 CI [0.110,0.348]), $\gamma$=  0.256 (0.68 CI [0.068,0.403], (0.95 CI [0.068,0.627]) }}
  \label{fig:SDSSfits}
  \end{figure}

The main advantage of this approach is that it not only does interpolation between data points, but also estimates the uncertainties in the interpolation. Figure \ref{fig:SDSSfits} shows two typical examples of SDSS S82 quasar light curves fitted by the stochastic process.

\clearpage
  \clearpage
  
Following \cite{Zu2011}, there are two important points to consider when comparing  these light curve reconstructions and the "error snakes" defined by the variances to the data points. 
First, these variances are the variances in the mean light curve and not the variance of the data relative to the mean light curve. The latter quantity is defined only when there is data and so it is ill-suited for showing a continuous light curve. Data points will be scattered relative to the mean light curve by the combination of the variance in the mean light curve and the uncertainties in the individual data points. Second, the reconstructed light curve is not an example of an individual light curve defined by the structure function parameters of the underlying model, but rather the average of all light curves defined by the structure function parameters that are consistent with the observed light curve given its uncertainties.

The variance in the reconstructed light curve is then the variance of these individual light curves about the mean. If we generated individual realizations of light curves each constrained by the data, they would track the mean light curve and statistically stay inside the "error snake" defined by the variances but they would show much more structure on short time scales and excursions outside the "error snakes" consistent with the estimated variances.

This approach can now easily be extended for the purpose of estimating a time delay $\tau_{\mathrm{delay}}$ between two (or more) light curves in order to estimate the size of the broad-line regions (BLRs) in AGNs. An approach for reverberation mapping based on this was first outlined by \cite{Zu2011}, and has been enhanced in this work to do broad-band reverberation mapping, supported by one epoch of spectroscopy to separate continuum and emission line contribution, and to use the power law instead of the DRW.

\section{REVERBERATION MAPPING COVARIANCE MATRIX ELEMENTS}
\label{sec:ReverberationMappingCovarianceMatrixElements}
In the case of a \textit{$\delta$-function transfer function}, one gets the following equations:

for the power law model:

i) covariance matrix for autocorrelation of the \textit{k} band flux: 
\begin{equation} \label{eqn:covxxfluxpowerlaw}  
 C_{ij}=C_{kk}^{cc}=\langle   f_k^c(t_i)  f_k^c(t_j)  \rangle =  A^2 \left[ \left( \frac{t_{\mathrm{obs}}}{1\; \mathrm{yr}}\right)^{\gamma} - \frac{1}{2} \left( \frac{|t_i-t_j|}{1 \; \mathrm{yr}}\right)^{\gamma} \; \right]
\end{equation}

ii) covariance matrix for the correlation function between \textit{k} band flux and \textit{l} band flux: 
\begin{equation} \label{eqn:covxyfluxpowerlaw}
\begin{split}
 C_{ij}= C_{kl}^{c,(e+c)}=\langle   f_k^c(t_i)  f_l^{ec}(t_j)  \rangle = &  s \, A^2 \left[ \left( \frac{t_{\mathrm{obs}}}{1 \; \mathrm{yr}} \right) ^{\gamma}- \frac{1}{2}  \left( \frac{|t_i-t_j|}{1 \;\mathrm{yr}}  \right)  ^{\gamma} \; \right] \\&+ e \, A^2 \left[ \left( \frac{t_{\mathrm{obs}}}{1 \; \mathrm{yr}} \right) ^{\gamma}- \frac{1}{2}  \left( \frac{|t_i-t_j+\tau_{\mathrm{delay}}|}{1 \;\mathrm{yr}}  \right)  ^{\gamma} \; \right]
\end{split}
\end{equation}

iii) covariance matrix for autocorrelation of the \textit{l} band flux: 
\begin{equation} \label{eqn:covyyfluxpowerlaw}
\begin{split}
 C_{ij}= C_{ll}^{(e+c),(e+c)}=\langle   f_l^{ec}(t_i)  f_l^{ec}(t_j)  \rangle =  & s^2 \, A^2 \left[ \left( \frac{t_{\mathrm{obs}}}{1 \; \mathrm{yr}} \right) ^{\gamma}- \frac{1}{2}  \left( \frac{|t_i-t_j|}{1 \; \mathrm{yr}}  \right)  ^{\gamma} \;  \right] \\
  & + s  e \, A^2 \left[ \left( \frac{t_{\mathrm{obs}}}{1 \; \mathrm{yr}} \right) ^{\gamma}- \frac{1}{2}  \left( \frac{|t_i-t_j+\tau_{\mathrm{delay}}|}{1 \; \mathrm{yr}}  \right)  ^{\gamma} \; \right]\\
 & + s e \, A^2 \left[ \left( \frac{t_{\mathrm{obs}}}{1 \; \mathrm{yr}} \right) ^{\gamma}- \frac{1}{2}  \left( \frac{|t_i-t_j-\tau_{\mathrm{delay}}|}{1 \; \mathrm{yr}}  \right)  ^{\gamma}\; \right]\\
& +e^2 A^2 \left[ \left( \frac{t_{\mathrm{obs}}}{1 \; \mathrm{yr}} \right) ^{\gamma}- \frac{1}{2}  \left( \frac{|t_i-t_j|}{1 \; \mathrm{yr}}  \right)  ^{\gamma}\; \right] 
\end{split}
\end{equation}
where $\tau_{\mathrm{delay}}$ is the time delay in years, $e$ is the line response of the \textit{l} band emission line to the flux in x band, and $s$ is the continuum response of the flux in $y$ band to the flux in \textit{k} band.\ \\

When using the DRW model instead, one gets the following equations:

i) covariance matrix for autocorrelation of the \textit{k} band continuum:  
\begin{equation}
 C_{ij}=\langle   f_k^c(t_i)  f_k^c(t_j)  \rangle = \frac{\omega^2}{2} \exp\left(- \frac{|t_i-t_j|}{\tau}   \right)
\end{equation}

ii) covariance matrix for the correlation function between \textit{k} band flux and \textit{l} band flux: 
\begin{equation} \label{eqn:covxyfluxdrw}
\begin{split}
 C_{ij}=\langle   f_k^c(t_i)  f_l^{ec}(t_j)  \rangle = &  s\, \frac{\omega^2}{2} \exp\left(- \frac{|t_i-t_j|}{\tau}   \right) \\&+ e \, \frac{\omega^2}{2} \exp\left(- \frac{|t_i-t_j + \tau_{\mathrm{delay}}|}{\tau}   \right)
\end{split}
\end{equation}

iii) covariance matrix for autocorrelation of the \textit{l} band flux: 
\begin{equation} \label{eqn:covyyfluxdrw}
\begin{split}
 C_{ij}=\langle   f_l^{ec}(t_i)  f_l^{ec}(t_j)  \rangle =  & s^2 \, \frac{\omega^2}{2} \exp\left(- \frac{|t_i-t_j|}{\tau}   \right) \\
  & + s \, e \cdot \frac{\omega^2}{2} \exp\left(- \frac{|t_i-t_j + \tau_{\mathrm{delay}}|}{\tau}   \right) \\ 
 & + s e \, \frac{\omega^2}{2} \exp\left(- \frac{|t_i-t_j - \tau_{\mathrm{delay}}|}{\tau}   \right) \\ 
& +e^2\,\frac{\omega^2}{2} \exp\left(- \frac{|t_i-t_j|}{\tau}   \right)
\end{split}
\end{equation}

In the case of using a \textit{Gaussian as transfer function}, we first carry out the covariance matrix of the cross-correlation between \textit{k} band continuum and \textit{l} band line

\begin{equation}
\begin{split}
\langle f_k^c(t_j)f_l^e(t_i) \rangle   &= \int _{t'=0}^{t_i} \mathrm{d}t' \Psi(t_i-t') \langle f_k^c(t')f_k^c(t_j) \rangle   \\
 &=  \int _{t'=0}^{t_i} \mathrm{d}t' \Psi_{\mathrm{Gauss}}(t_i-t') \langle f_k^c(t')f_k^c(t_j) \rangle   \\
 &=  \int _{t'=0}^{t_i} \mathrm{d}t' e \,\frac{1}{\sqrt{2 \pi \sigma_{\mathrm{Gauss}}^2}} \exp\left[ -\frac{(t_i-t'-\tau_{\mathrm{delay}})^2}{2\sigma_{\mathrm{Gauss}}^2}    \right] \langle f_k^c(t')f_k^c(t_j) \rangle   \\
 &= e \, \frac{1}{\sqrt{2 \pi \sigma_{\mathrm{Gauss}}^2}} \int _{t'=0}^{t_i} \mathrm{d}t'  \exp\left[ -\frac{(t_i-t'-\tau_{\mathrm{delay}})^2}{2\sigma_{\mathrm{Gauss}}^2}    \right] \langle f_k^c(t')f_k^c(t_j) \rangle  
\end{split}
\end{equation}

where $\mathrm{erf}(z)= \frac{2}{\sqrt{\pi}} \int_0^z \exp \left[ -t^2   \right] \mathrm{d}t \stackrel{\mathrm{McLaurin \; series}}{=} \frac{2}{\pi} (z- \frac{z^3}{3} + \frac{z^5}{10}  - \frac{z^7}{42} + \frac{z^9}{216} - ...)  $. This can be calculated with the \texttt{gsl} function \texttt{double gsl\_sf\_erf(double x)}.\ \\

Also, we carry out the covariance matrix of the autocorrelation of the \textit{l} band line

\begin{small}
\begin{equation}
\begin{split}
\langle f_l^e(t_j)f_l^e(t_i) \rangle   &= \int _{t'=0}^{t_i} \mathrm{d}t' \int _{t''=0}^{t_j} dt'' \Psi(t_i-t') \Psi(t_j-t'') \langle f_k^c(t')f_k^c(t'') \rangle   \\
 &= \int _{t'=0}^{t_i} \mathrm{d}t' \int _{t''=0}^{t_j} dt'' \Psi_{\mathrm{Gauss}}(t_i-t') \Psi_{\mathrm{Gauss}}(t_j-t'') \langle f_k^c(t')f_k^c(t'') \rangle   \\
 &= \int _{t'=0}^{t_i} \mathrm{d}t' \int _{t''=0}^{t_j} dt'' \\
  & \cdot \frac{e^2}{ 2 \pi \sigma_{\mathrm{Gauss}}^2} \cdot \exp \left[ -\frac{(t_i-t'-\tau_{\mathrm{delay}})^2}{2\sigma_{\mathrm{Gauss}}^2}     \right]   \, \exp \left[ -\frac{(t_j-t''-\tau_{\mathrm{delay}})^2}{2\sigma_{\mathrm{Gauss}}^2}     \right]    \langle f_x^c(t')f_x^c(t'') \rangle   \\
\end{split}
\end{equation}
\end{small}

For the power law model, 

\begin{equation}
\begin{split}
\langle f_k^c(t_j)f_l^e(t_i) \rangle   &= e \,\frac{1}{\sqrt{2 \pi \sigma_{\mathrm{Gauss}}^2}} \int _{t'=0}^{t_i} \mathrm{d}t'  \exp\left[ -\frac{(t_i-t'-\tau_{\mathrm{delay}})^2}{2\sigma_{\mathrm{Gauss}}^2}    \right] \langle f_x^c(t')f_x^c(t_j) \rangle  \\
&=e \, \frac{1}{\sqrt{2 \pi \sigma_{\mathrm{Gauss}}^2}} \int _{t'=0}^{t_i} \mathrm{d}t'  \exp\left[ -\frac{(t_i-t'-\tau_{\mathrm{delay}})^2}{2\sigma_{\mathrm{Gauss}}^2}    \right]  A^2 \left(        t_{\mathrm{obs}}^{\gamma} - \frac{1}{2} \left(    \frac{|t_j-t'|}{1 \; \mathrm{yr}}  \right)^{\gamma} \; \right)
\end{split}
\end{equation}

\begin{small}
\begin{equation}
\begin{split}
\langle f_l^e(t_j)f_l^e(t_i) \rangle   &= \int _{t'=0}^{t_i} \mathrm{d}t' \int _{t''=0}^{t_j} \mathrm{d}t''\, \frac{e}{ \sqrt{2 \pi \sigma_{\mathrm{Gauss}}^2}} \, \exp \left[ -\frac{(t_i-t'-\tau_{\mathrm{delay}})^2}{2\sigma_{\mathrm{Gauss}}^2}     \right]  \\ & \quad  \quad \cdot \frac{e}{ \sqrt{2 \pi \sigma_{Gauss}^2}} \, \exp \left[ -\frac{(t_j-t''-\tau_{\mathrm{delay}})^2}{2\sigma_{\mathrm{Gauss}}^2}     \right]      \langle f_x^c(t')f_k^c(t'') \rangle   \\
&= \int _{t'=0}^{t_i} \mathrm{d}t' \int _{t''=0}^{t_j} \mathrm{d}t''\,\frac{e^2}{ 2 \pi \sigma_{\mathrm{Gauss}}^2} \\
& \quad \quad\cdot \exp \left[ -\frac{(t_i-t'-\tau_{\mathrm{delay}})^2}{2\sigma_{\mathrm{Gauss}}^2}     \right] \, \exp \left[ -\frac{(t_j-t''-\tau_{\mathrm{delay}})^2}{2\sigma_{\mathrm{Gauss}}^2}     \right] \\ & \quad\quad \cdot A^2 \left( t_{\mathrm{obs}}^{\gamma} - \frac{1}{2} \left(    \frac{|t_j-t'|}{1 \; \mathrm{yr}}  \right)^{\gamma} \; \right)
\end{split}
\end{equation}
\end{small}

Unfortunately, these integrals for a power law structure function are not analytical.\ \\

For using a DRW structure function instead, the integrals are analytical:

covariance matrix of the cross-correlation between \textit{k} band continuum and \textit{l} band line
\begin{small}
\begin{equation}
\begin{split}
\langle f_k^c(t_j)f_l^e(t_i) \rangle   &= e \, \frac{1}{\sqrt{2 \pi \sigma_{\mathrm{Gauss}}^2}} \int _{t'=0}^{t_i} \mathrm{d}t'  \exp\left[ -\frac{(t_i-t'-\tau_{\mathrm{delay}})^2}{2\sigma_{\mathrm{Gauss}}^2}    \right] \langle f_k^c(t')f_k^c(t_j) \rangle  \\
&=e \, \frac{1}{\sqrt{2 \pi \sigma_{\mathrm{Gauss}}^2}} \int _{t'=0}^{t_i} \mathrm{d}t'  \exp\left[ -\frac{(t_i-t'-\tau_{\mathrm{delay}})^2}{2\sigma_{\mathrm{Gauss}}^2}    \right]   \hat{\omega}^ 2 \exp \left[ - \frac{|t_j - t'|}{\tau}    \right] \\
&= \frac{1}{2} \, \exp \left[ \frac{\sigma_{\mathrm{Gauss}}^ 2 - 2k\tau (\tau_{\mathrm{delay}}+t_j-t_i)      }{2 \tau ^2}      \right]\, \hat{\omega}^ 2    \sigma_{\mathrm{Gauss}} \\ & \quad  \cdot e \left(\mathrm{erf}\left[ \frac{\tau_{\mathrm{delay}}\tau - w \sigma_{\mathrm{Gauss}}^ 2 }{\sqrt{2}\tau \sigma_{Gauss}^2}\right]  -
\mathrm{erf}\left[ \frac{\tau_{\mathrm{delay}}\tau - w \sigma_{\mathrm{Gauss}}^ 2 - \tau t_i}{\sqrt{2}\tau \sigma_{\mathrm{Gauss}}^2}   \right]     \right)
\end{split}
\label{eqn:gausscorrxy}
\end{equation}
\end{small}
with
\begin{equation}
\hat{\omega}= \sqrt{\frac{2 \omega ^2}{\tau}}
\end{equation}
and
\begin{equation}
w=\left\{\begin{array}{cl} +1, & \mbox{if } t_j>t_i   \\ -1, & \mbox{if } t_j < t_i\end{array}\right.
\end{equation}
where per definition $t_j \neq t_i$.

covariance matrix of the autocorrelation of the \textit{l} band line
\begin{small}
\begin{equation}
\begin{split}
\langle f_l^e(t_j)f_l^e(t_i) \rangle   &= \int _{t'=0}^{t_i} \mathrm{d}t' \int _{t''=0}^{t_j} \mathrm{d}t''\, \frac{e}{ \sqrt{2 \pi \sigma_{\mathrm{Gauss}}^2}} \, \exp \left[ -\frac{(t_i-t'-\tau_{\mathrm{delay}})^2}{2\sigma_{\mathrm{Gauss}}^2}     \right]  \\ & \quad \quad   \cdot \frac{e}{ \sqrt{2 \pi \sigma_{\mathrm{Gauss}}^2}} \, \exp \left[ -\frac{(t_j-t''-\tau_{\mathrm{delay}})^2}{2\sigma_{\mathrm{Gauss}}^2}     \right]                \langle f_k^c(t')f_k^c(t'') \rangle   \\
&= \int _{t'=0}^{t_i} \mathrm{d}t' \int _{t''=0}^{t_j} \mathrm{d}t''\, \frac{e}{ \sqrt{2 \pi \sigma_{\mathrm{Gauss}}^2}} \, \exp \left[ -\frac{(t_i-t'-\tau_{\mathrm{Gauss}})^2}{2\sigma_{\mathrm{Gauss}}^2}     \right]  \\ &\quad \quad \cdot \frac{e}{ \sqrt{2 \pi \sigma_{\mathrm{Gauss}}^2}} \cdot \exp \left[ -\frac{(t_j-t''-\tau_{\mathrm{delay}})^2}{2\sigma_{\mathrm{Gauss}}^2}     \right]   \hat{\omega}^ 2 \exp \left[ - \frac{|t_t' - t''|}{\tau}    \right] \\
&= \frac{1}{4 } \cdot \exp \left[ - \frac{w t_j \tau^ 2 + w t_i \tau}{\tau^ 2}       \right] \hat{\omega}^ 2  e^ 2 \\   & \quad \cdot \left( \mathrm{erf}\left[ \frac{w\tau_{\mathrm{delay}} \tau + \sigma_{\mathrm{Gauss}}^ 2 }{\sqrt{2}\tau^2 \sigma_{\mathrm{Gauss}}^ 2}  \right] -  \mathrm{erf}\left[ \frac{w\tau_{\mathrm{delay}} \tau - w t_j \tau + \sigma_{\mathrm{Gauss}}^ 2}{\sqrt{2} \tau^2 \sigma_{\mathrm{Gauss}}^ 2 }  \right]        \right) \\ & \quad \cdot   \left(  \mathrm{erf}\left[ \frac{\tau_{\mathrm{delay}} \tau - w \sigma^2_{\mathrm{Gauss}}}{\sqrt{2} \tau \sigma^ 2_{\mathrm{Gauss}}}      \right] -   \mathrm{erf}\left[ \frac{\tau_{\mathrm{delay}} \tau - w \sigma^2_{\mathrm{Gauss}} - \tau t_i}{\sqrt{2} \tau \sigma^ 2_{\mathrm{Gauss}}}      \right]     \right) 
\end{split}
\label{eqn:gaussautocorry}
\end{equation}
\end{small}

with
\begin{equation}
w=\left\{\begin{array}{cl} +1, & \mbox{if } t_j>t_i   \\ -1, & \mbox{if } t_j < t_i\end{array}\right.
\end{equation}
where per definition $t_j \neq t_i$.

From this, we can calculate the covariance matrix for the correlation function between \textit{k} band flux and \textit{l} band flux and the covariance matrix for autocorrelation of the \textit{l} band flux by inserting \eqref{eqn:gausscorrxy} and \eqref{eqn:gaussautocorry} into \eqref{eqn:cfxfy} and \eqref{eqn:cfyfy}.

\section{TEST DATA}
\label{sec:TestData}

Equipped with a statistical description of quasar variability (see Section \ref{sec:QuasarLightCurvesAsAStochasticProcess}), we generate well-sampled mock light curves in order to (i) test the algorithm for determination of the structure function parameters and later for reverberation mapping, (ii) demonstrate the relationship between our model parameters and the shape of light curves, (iii) estimate the systematic effects that sampling rate and light curve length have on the fitted parameter both for determination of the structure function parameter and for reverberation mapping.
The later is especially important because the SDSS S82 data are fairly sparse.

\subsection{Generating Test Data}
\label{sec:GeneratingTestData}

A continuum light curve is generated using only the two structure function parameters and the mean magnitude of the light curve as input parameters. To apply a time delay $\tau_{\mathrm{delay}}$ to continuum plus emission line light curves, we refer to our assumption that all emission line light curves are scaled, smoothed and displaced (delayed) versions of the continuum flux light curve, see Equ. \eqref{eqn:fxfymodela}. 

For generating the test data, one can choose
\begin{itemize}
\item the structure function; here, a DRW model and a power-law model are implemented
\item the transfer function $\Psi(\tau_{\mathrm{delay}})$; here, a $\delta$ function and a Gaussian are implemented
\item scaling factors $s$, $e$ (instead of emission line equivalent width)
\item the bands $x$ and $y$, also, multiple emission lines are possible
\item the time sampling and time windows; this can be chosen freely and also be inherited from SDSS S82 light curves.
\end{itemize}
As mock data are generated from $e$ and $s$, and not with emission line equivalent width, we replace \eqref{eqn:likspek} by
\begin{equation}
P(e)=\frac{1}{\sqrt{2 \pi} \delta e_0}\exp \left( - \frac{(e-e_0)^ 2}{2 \delta e0^ 2}   \right),
\label{eqn:mockspec}
\end{equation}
when testing with mock data. The values for $e_0$ and $\delta e_0$ are set depending on test, typical $e_0 \approx 0.2$, $\delta e_0 \approx 0.02$.

\begin{figure}[H]
			 \subfigure[ ]        {
            \includegraphics{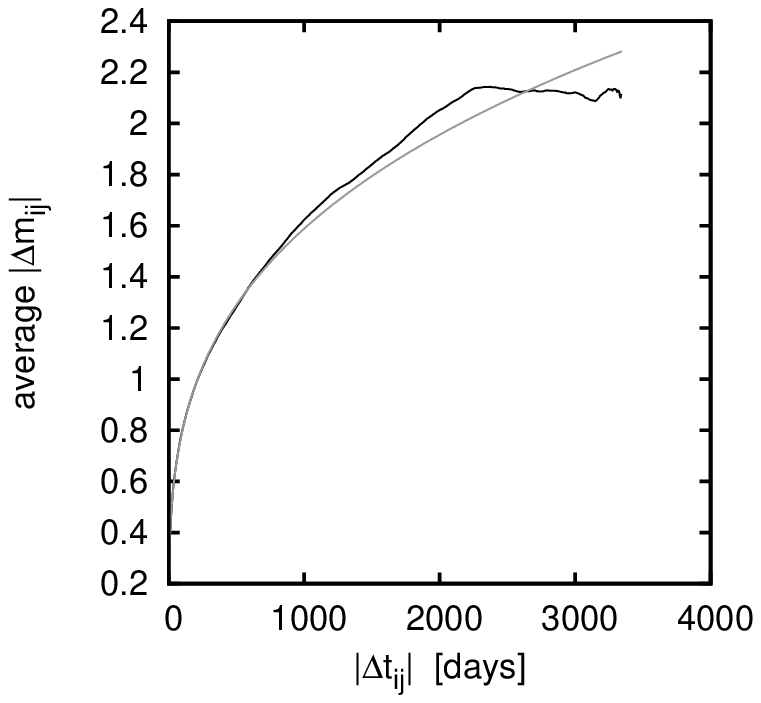}      \qquad
             }      
			 \subfigure[ ]
            {
              \includegraphics{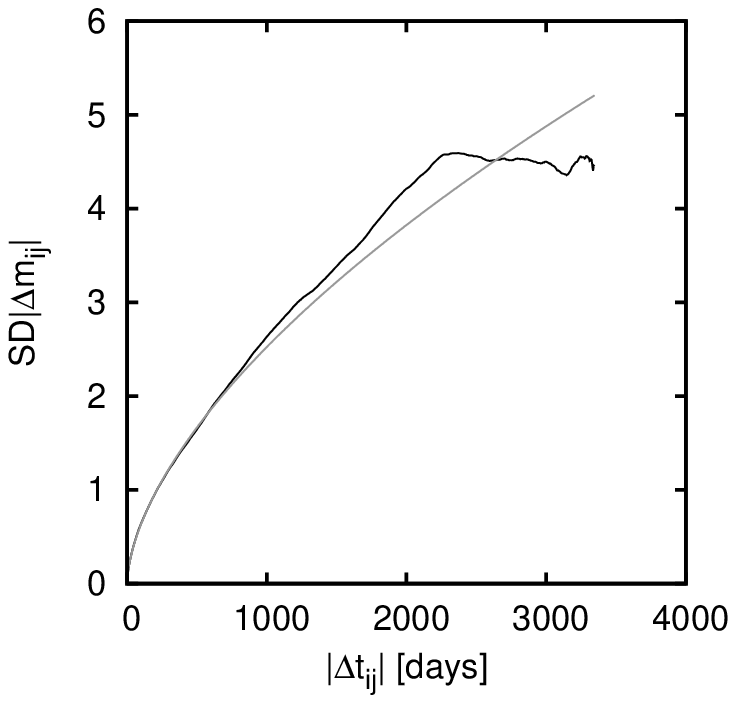}   
            }
\caption[light curve statistics]{\footnotesize {light curve statistics over $|\Delta t_{ij}|$ for 100 simulated power law light curves, where $i$,$j$ are the individual observations; \newline (a) black curve: averaged $|\Delta m_{ij}|$, grey curve: $\sqrt{V(|\Delta t|)}$, \newline (b) black curve: standard deviation of $|\Delta m_{ij}|$, grey curve: $V(|\Delta t|)$}}
\label{fig:average_dmdt_standarddev}
\end{figure}

To make sure that the test data are consistent, samples of mock light curves are evaluated statistically.\newline
When averaging over a sample of 100 light curves having the same structure function $\mathrm{V}$, the form of the point cloud $|\Delta m|$ vs. $|\Delta t|$ should be  represented by $\sqrt{\mathrm{V}(| \Delta t |)}$ as a ridge-line. Also, the standard deviation $\mathrm{SD}(|\Delta m|)$  should be represented by the structure function itself, $\mathrm{V}(| \Delta t |)$. \newline
For the purpose of illustrating the effects, an observation time of 12 years, observational time sampling of 1 day, was simulated. The first and last year of the data points are neglected to avoid potential edge effects, resulting in effective 10 years observational time. As shown in Fig. \ref{fig:average_dmdt_standarddev}, for both the power law and the damped random walk structure functions, the form of the point cloud $|\Delta m|$ vs. $|\Delta t|$ is represented quite well by $\sqrt{\mathrm{V}(| \Delta t |)}$ as a ridge-line and the standard deviation is represented by the structure function itself, $\mathrm{V}(| \Delta t |)$. Also, it looks plausible that about 63 percent of points of $|\Delta m|$ vs. $|\Delta t|$ will be under the line (fraction of data being within 1 $\sigma$ in a Gaussian distribution). As a result, we know that the generated mock light curves follow our assumptions about structure functions. Because of this, we can be sure that such mock light curves can be used safely to test our algorithms for estimating structure function parameters and reverberation mapping.

\clearpage

\subsection{Results from Test Data} %
\label{sec:ResultsFromTestData}

In this section, we show results from mock data, that lead us to an understanding of the behavior of the reverberation mapping algorithm.

Estimation of the structure function parameter is done by evaluating Equ. \eqref{eqn:parameterlikelihood} as described in Chapter \ref{sec:QuasarLightCurvesAsAStochasticProcess}. This can be carried out on a parameter grid or by using a MCMC method.
Using a parameter grid is not practicable as it is time consuming, but it is functional for testing issues and to show the likelihood surface.

We found that the method is sufficient for estimating structure function parameters and fitting light curves, as even for sparsely sampled mock data it is able to
recover the input values within about a 68 \% confidence interval or better. Also we see that the posterior distribution in the parameter space, like the likelihood surface itself, is smooth, so the MCMC is able to sample the posterior distribution.

\begin{figure}[H]
			 \subfigure[ ]        {
            \includegraphics{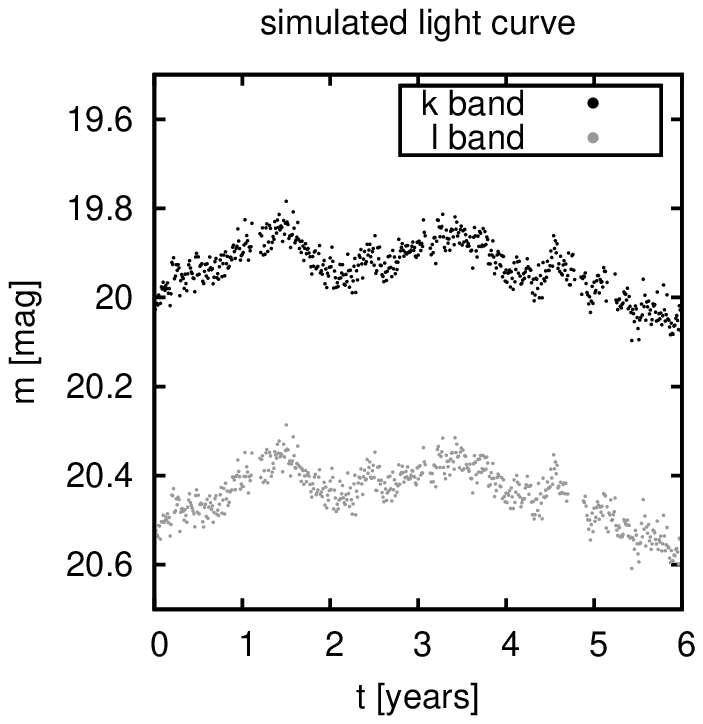}      \qquad
             }      
			 \subfigure[ ]
            {
              \includegraphics{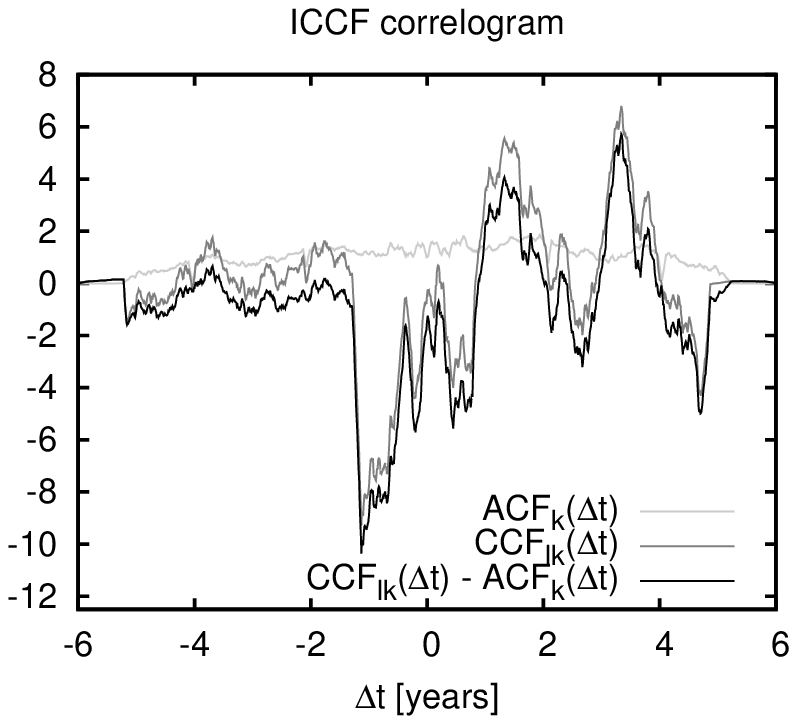}   
            }
\caption[ICCF correlogram]{\footnotesize{Non-uniform sampled simulated light curve and its correlogram, input $\tau_{\mathrm{delay}}$ = 1 year is recovered as  $\tau_{\mathrm{delay}}$ = 3.33 years. Band $k$ contains continuum only, band $l$ contains continuum and emission line.}}
\label{fig:iccf}
\end{figure}

During first tests on estimation of the time delay $\tau_{\mathrm{delay}}$, uniformly sampled mock data were used. Using uniformly sampled light curves spanning a time longer than the input $\tau_{\mathrm{delay}}$, this delay can be recovered with a precision of 1 over the sampling frequency. This holds even when the flux contribution is about 5 percent (the value that is expected for many of the SDSS S82 light curves, whereas some have a flux contribution up to 20 percent). For more realistic tests, we have taken typical SDSS S82 time sampling and applied it to mock light curves. As expected, the approach is sensitive to time sampling, but not as much as "classical" approaches like e.g. CCF and ICCF. For an example plot, see Fig. \ref{fig:iccf}.

We have found that we can easily estimate if a given time sampling enables us to find a time delay in an expected range. The tool used for this is the histogram of observational time lags provided by the light curve in question. Details on this can be found in Section \ref{sec:ApplicationToSDSSS82Data}.

Despite the robustness of the stochastic approach, results must be handled with care. Even when the algorithm is able to recover the input $\tau_{\mathrm{delay}}$, we found that for sparsely sampled data the posterior distribution often turns out to be much more flat than for estimating the structure function parameters. Additionally, periodic pattern of higher and lower likelihood can be found in mock and real data. This pattern is reproducing roughly the pattern in the histogram of observational time lags. As a result of this, we constrain the $\tau_{\mathrm{delay}}$ by \eqref{eqn:priorgauss} and set a prior on $(e,s)$ based on spectroscopic data. For details, see Section \ref{sec:ApplicationToSDSSS82Data}, Equ. \eqref{eqn:intfore} to \eqref{eqn:likspek}.\newline
If those priors are applied, this periodic pattern appears only in some cases. When it appears, it is only weak. Most posterior distribution functions are smooth and roundish, others have a stretched appearance or show a "tail" which is a remnant of the periodic pattern. This demonstrates the importancy of choosing a sensible prior.

%% file: Tables.tex
\section*{TABLES}
\label{cha:Tables}
\begin{deluxetable}{lllll}
\tabletypesize{\scriptsize}
\tablecaption{Emission Lines\label{tab:linecontribution}}
\tablewidth{0pt}
\tablehead{
\colhead{redshift} & \colhead{\textit{g}} & \colhead{\textit{r}} & \colhead{\textit{i}} & \colhead{\textit{z}}
}
\startdata
0.08	 &	 H$\beta$, H$\gamma$ (plus some other)&	continuum &	-&	 continuum \\
0.13 - 0.142 	& continuum &	 continuum &	 H$\alpha$ (plus some other)	& continuum\\
0.225 - 0.283 & continuum 	& H$\beta$, H$\gamma$ (plus some other) &	 continuum &	 continuum \\
0.284 - 0.291	&	continuum &	 H$\beta$, H$\gamma$ (\ionscale{O}{iii} $\lambda$4959, $\lambda$5007)	& continuum &	- \\			
0.349 - 0.371 	&  continuum 	& continuum &	 continuum &	 H$\alpha$ (plus some other)\\					
0.463	& continuum &	 continuum &	-	& H$\alpha$ (plus some other)\\			
0.519 - 0.537	 	& continuum &	 continuum &	 H$\beta$, H$\gamma$ (plus some other)	& continuum \\			
0.538 - 0.552  &	 \ionscale{Mg}{ii} &	 continuum 	& H$\beta$, H$\gamma$ (\ionscale{O}{iii} $\lambda$4959, $\lambda$5007) &  continuum \\
0.553 - 0.554  	&  \ionscale{Mg}{ii} &	 continuum &	-	& continuum \\
0.555 - 0.591 &	  \ionscale{Mg}{ii} 	& continuum 	& continuum 	& continuum \\
0.592 - 0.732  &	  \ionscale{Mg}{ii} (\ionscale{Fe}{ii} $\lambda$2600)	& continuum 	& continuum &	 continuum \\
0.733 - 0.813		&  \ionscale{Mg}{ii} (\ionscale{Fe}{ii} $\lambda$2600)	& continuum 	& continuum &	- \\			
0.814 - 0.846  &  \ionscale{Mg}{ii} (\ionscale{Fe}{ii} $\lambda$2600) & continuum &	continuum &	 H$\beta$, H$\gamma$ (plus some other) \\			
0.847 - 0.851  & 	-	 & continuum & 	 continuum &	 H$\beta$, H$\gamma$ (\ionscale{O}{iii} $\lambda$4959, $\lambda$5007) \\			
1.171 - 1.191 	&	 continuum &	  \ionscale{Mg}{ii} &	 continuum & 	 continuum \\
1.192 - 1.207 	& continuum &	  \ionscale{Mg}{ii} &	 continuum &	 continuum \\
1.765 - 1.786 	& \ionscale{C}{vi} &	 continuum &	 continuum 	& continuum \\
1.787 - 1.912 	& \ionscale{C}{vi} &	 continuum & 	-	& continuum \\
1.913 - 2.036  &	 \ionscale{C}{vi} &	 continuum &	 continuum &	 continuum \\
2.037 - 2.185 	& \ionscale{C}{vi} (\ionscale{Si}{vi})	& continuum &	 continuum & 	 continuum \\
2.186 - 2.254 	& \ionscale{C}{vi} (\ionscale{Si}{vi})	& -& 	 continuum 	& continuum \\
2.255 - 2.32 	& \ionscale{C}{vi} (\ionscale{Si}{vi})	& continuum &	 continuum &	 continuum \\
2.903 - 2.969  &	-	& \ionscale{C}{vi} &	 continuum 	& continuum \\  
\enddata
\tablecomments{In a compact form this table lists which emission lines can be used for a given redshift and in which cases there is more than one emission line in a band. \newline
"-" indicates this band cannot be used (a continuum and some line contribution from a line that is not used as emission line for reverberation mapping). \newline
Sometimes there is a emission line that can be used but also (weak) contribution from another lines, those are written in brackets.}
\end{deluxetable}

\clearpage
\LongTables
. Make sure there is at least one \tablenotemark
\tablenotetext{a}{\texttt{headobjid} from SDSS tables}
\tablenotetext{b}{redshift}
\tablenotetext{c}{restframe equivalent width of broad line}
\tablenotetext{d}{FWHM of broad line}
\tablenotetext{e}{computed from Kaspi relation, in rest frame}
\tablenotetext{f}{own computation from all points within the 68 \% CI, in rest frame}
\end{deluxetable*}
\clearpage